\documentclass[aps,prb,preprint,amsmath,amssymb,superscriptaddress]{revtex4-2}
\usepackage[dvipsnames]{xcolor}
\usepackage{graphicx}
\usepackage{caption}
\usepackage{ragged2e}
\usepackage{hyperref}
\usepackage{subcaption} 

\makeatletter
\renewcommand\p@subfigure{} 
\makeatother
\usepackage{dcolumn}
\usepackage{bm}
\usepackage[normalem]{ulem}
\usepackage[utf8]{inputenc}
\usepackage[T1]{fontenc}
\usepackage{mathptmx}
\usepackage{etoolbox}
\usepackage{booktabs}
\definecolor{azure(colorwheel)}{rgb}{0.0, 0.5, 1.0}
\makeatother

\usepackage{booktabs}

\begin{document}

\title{Long-lived relaxation channel and exciton-phonon coupling in \textrm{Ta\textsubscript{2}NiSe\textsubscript{5}} via nondegenerate pump-probe spectroscopy}

\author{Poulami Ghosh}
\affiliation{Department of Physical Sciences, Indian Institute of Science Education and Research Kolkata, Mohanpur 741246, West Bengal, India.}

\author{Anupama Chauhan}
\affiliation{Department of Physical Sciences, Indian Institute of Science Education and Research Kolkata, Mohanpur 741246, West Bengal, India.}

\author{Sidhanta Sahu}
\affiliation{Department of Physical Sciences, Indian Institute of Science Education and Research Kolkata, Mohanpur 741246, West Bengal, India.}

\author{Sk Kalimuddin}
\affiliation{School of Physical Sciences, Indian Association for the Cultivation of Science, 2A \& 2B Raja S C Mullick Road, Jadavpur, Kolkata 700032, India.}

\author{Mintu Mondal}
\affiliation{School of Physical Sciences, Indian Association for the Cultivation of Science, 2A \& 2B Raja S C Mullick Road, Jadavpur, Kolkata 700032, India.}

\author{N. Kamaraju}
\email{nkamaraju@iiserkol.ac.in}
\affiliation{Department of Physical Sciences, Indian Institute of Science Education and Research Kolkata, Mohanpur 741246, West Bengal, India.}
\altaffiliation{Department of Physical Sciences, Indian Institute of Science Education and Research Kolkata, Mohanpur 741246, West Bengal, India.}

\date{\today}

\begin{abstract}
An excitonic insulator represents a quantum phase in which spontaneous condensation of excitons leads to novel many-body phenomena. Ta$_2$NiSe$_5$ (TNSe), a layered narrow-gap semiconductor, has emerged as a model platform to probe these correlated excitonic phases and their underlying dynamics below 327~K. In this work, we investigate the nonequilibrium dynamics of TNSe using temperature-dependent, non-degenerate optical pump-probe spectroscopy with a 3.14~eV pump and a 1.57~eV probe, extending the accessible pump-probe delay window up to 500 ps. In addition to the well-established subpicosecond relaxation channel ($\sim$0.7-0.9~ps) associated with carrier cooling and recombination, accompanied by exciton reformation, we uncover a much slower recovery process with a decay time of $\sim$280-600~ps, significantly longer than previously reported. We attribute this unusually prolonged recovery to enhanced scattering between excitons and nonequilibrium phonons, which delays the reestablishment of equilibrium excitonic correlations. On top of this biexponential background, we observe two coherent phonon modes at 1.0 and 2.9~THz with distinctly different coupling behaviors. The 1.0~THz mode exhibits an order-parameter-like temperature dependence, consistent with strong coupling to the excitonic condensate in TNSe. In contrast, the 2.9~THz mode does not exhibit any discernible coupling to the excitonic order parameter, and appears to arise from anharmonic lattice dynamics associated with the structural phase transition. Together, these results elucidate the hierarchy of relaxation pathways in TNSe and highlight the importance of extending the temporal detection window in pump-probe measurements to fully capture long-lived exciton-phonon dynamics.
\end{abstract}


\maketitle 

\section{\label{sec:level1}Introduction}

\begin{figure*}
    \centering
    \includegraphics[width=\textwidth]{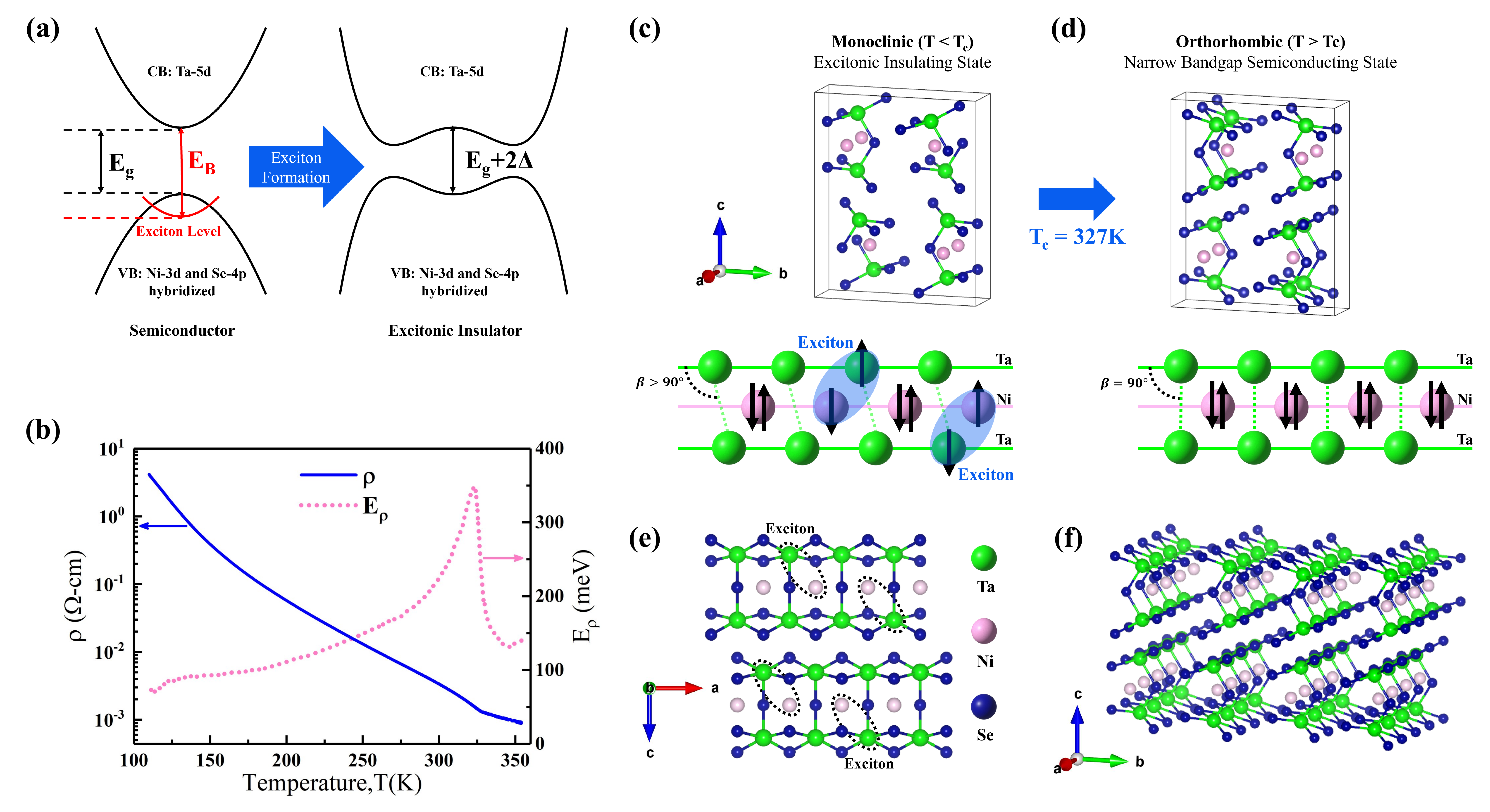}
    \begin{subfigure}{0\linewidth}
        \phantomcaption
        \label{fig:1a}
    \end{subfigure}
    \begin{subfigure}{0\linewidth}
        \phantomcaption
        \label{fig:1b}
    \end{subfigure}
    \begin{subfigure}{0\linewidth}
        \phantomcaption
        \label{fig:1c}
    \end{subfigure}
    \begin{subfigure}{0\linewidth}
        \phantomcaption
        \label{fig:1d}
    \end{subfigure}
    \begin{subfigure}{0\linewidth}
        \phantomcaption
        \label{fig:1e}
    \end{subfigure}
    \begin{subfigure}{0\linewidth}
        \phantomcaption
        \label{fig:1f}
    \end{subfigure}
    \caption{\justifying (a) Schematic band diagram illustrating the transition into an excitonic insulator (EI) state. In a narrow-gap semiconductor, the EI phase emerges when the exciton binding energy (E\textsubscript{B}) exceeds the band gap (E\textsubscript{g}), leading to pronounced band renormalization near the Fermi level. (b) Temperature variation of resistivity, $\rho$ (blue, left axis) (the current, \textit{I}, is passed $\parallel$ to the \textit{a} axis) and the calculated temperature variation of activation energy, $E_\rho$  (pink, right axis). The schematic crystal structures of TNSe in the (c) low-temperature and (d) high-temperature phases. (e) and (f) denote the crystal structures viewed along the \textit{b} and \textit{a} axis, respectively. 1D chains of Ta (green) and Ni (pink) run along the \textit{a} axis, forming layered sheets in the ac plane, with Se atoms depicted in blue.}
    \label{fig:1}
\end{figure*}

The spontaneous formation and condensation of bound electron-hole pairs (excitons) can drive a system into an insulating ground state, realizing the excitonic insulator phase—a quantum phase of matter \cite{kohn1967excitonic, jerome1967excitonic, halperin1968possible}. Such a phase may arise in both narrow-gap semiconductors and semimetals, depending on the balance between the excitonic binding energy and band structure \cite{bronold2006possibility,zenker2012electron,keldysh2024possible}. In narrow-gap semiconductors, when the exciton binding energy exceeds the band gap [as seen in Fig.\ref{fig:1a}], the system can enter into an excitonic insulator phase via condensation of preformed excitons, akin to Bose-Einstein condensation (BEC) \cite{knox1963theory}. On the other hand, in certain semimetallic systems, reduced carrier screening can allow exciton formation and condensation to occur simultaneously at the transition, similar to BCS-like pairing \cite{gao2023evidence}. In either case, the resulting EI state is marked by flattening of bands near the Fermi level, with the excitonic binding energy determining the low-temperature gap.

Over the years, various classes of materials \cite{bucher1991excitonic,monney2009spontaneous,kunevs2014excitonic,kogar2017signatures,wang2018excitonic,varsano2020monolayer,huang2024evidence} have been proposed as potential candidates for hosting the EI phase.  Among these, Ta\textsubscript{2}NiSe\textsubscript{5} (TNSe) stands out due to its quasi-1D crystal structure, stability under ambient conditions, and reportedly a small direct band gap of 0.36 eV \cite{sunshine1985structure}, placing it close to the boundary between semiconducting and semimetallic regimes. These properties make TNSe a compelling platform for investigating EI physics in a real condensed matter system. However, despite extensive experimental \cite{wakisaka2009excitonic,mor2017ultrafast,werdehausen2018photo,werdehausen2018coherent,yan2019strong,liu2021photoinduced,lu2021evolution,baldini2023spontaneous,okamura2023pressure,chen2025structural} and theoretical \cite{kaneko2013orthorhombic,murakami2017photoinduced,subedi2020orthorhombic,mazza2020nature,chatterjee2025collective} studies, the nature of the insulating phase in TNSe remains under active debate, largely due to the concomitant structural transition and gap opening at the critical temperature, ($T_c$). This has raised the central question of whether the low-temperature gap is primarily driven by excitonic correlations or by electron-lattice interactions. Baldini \textit{et al.} \cite{baldini2023spontaneous} used time-resolved optical spectroscopy combined with x-ray scattering to argue that the phase transition in TNSe is primarily a lattice-driven structural distortion rather than a purely excitonic condensate. However, both prior \cite{yan2019strong,kim2021direct,miyamoto2022charge} and more recent studies \cite{lee2024surface,arora2025pressure,tang2025non,bae2025microscopic,shi2025uniaxial,rosenberg2026excitonic} have reported persistent excitonic signatures, tunable electron-hole correlations, and strong coupling between electronic and lattice degrees of freedom, challenging the purely structural interpretation and suggesting that the insulating gap likely results from a combination of lattice and electronic effects.

Previous ultrafast studies have explored the coherent phonon oscillations, characteristic relaxation timescales associated with gap dynamics, and signatures of excitonic correlations across a wide range of temperatures and excitation fluences \cite{mor2017ultrafast,werdehausen2018photo,werdehausen2018coherent,liu2021photoinduced,miyamoto2022charge,jiang2024multiple,arora2025ultrafast,bretscher2021ultrafast}. However, most of the investigations have employed excitation energies at or below $\sim1.55$~eV \cite{werdehausen2018photo,bretscher2021ultrafast,jiang2024multiple,arora2025ultrafast}, primarily probing carrier dynamics near the band edge. In addition, the accessible pump-probe delay window in many of these studies is typically limited to a few tens of picoseconds, potentially overlooking slower relaxation channels. Exciting with higher-energy photons, such as 3.14~eV pump pulses, allows direct access to higher-lying electronic states and hot carriers. This approach enables us to access relaxation pathways beyond the near-band-edge regime and to probe their coupling to lattice degrees of freedom over extended timescales. In this work, we combine high-energy optical excitation with an extended pump-probe delay window of up to 500 ps to probe temperature-dependent nonequilibrium carrier-phonon dynamics in TNSe across the coupled excitonic-structural phase transition.

\section{\label{sec:level2}Experimental Details}
\subsection{\label{sec:levelA}Sample growth and characterization}

Single crystals of Ta$_2$NiSe$_5$ are synthesized using the chemical vapor transport method. The details of the growth process is provided in the Supplemental Material (SM) \cite{supplemental} (see also Refs. \cite{katsumi2023disentangling,laidler1984arrhenius,zhang2021thermal,Villars2023crystal,haque2024terahertz,golez2022unveiling} therein) Sec. S1. The temperature-dependent resistivity measurements of the as-grown crystals [Fig.\ref{fig:1b}] exhibit a clear anomaly near ($T_c \approx 327$)~K, consistent with the concomitant excitonic-insulator and structural phase transition in TNSe \cite{sunshine1985structure,di1986physical,yan2019strong,lu2017zero}. To further understand the transport mechanism, we analyze the resistivity in terms of activation model\cite{neifel2007temperature}, as discussed in the SM \cite{supplemental} (Sec. S1.3).

The single crystal of TNSe has a layered, quasi-1D structure composed of parallel chains of Ta-Se octahedra and Ni-Se tetrahedra running along the crystallographic \textit{a}-axis \cite{sunshine1985structure}, as shown schematically in Figs.\ref{fig:1c}-\ref{fig:1f}. Above $T_c$, TNSe crystallizes in the orthorhombic phase (space group $Cmcm$) and behaves as a narrow-gap semiconductor, whereas on cooling it undergoes a semiconductor-to-insulator transition accompanied by a structural phase transition from orthorhombic (\textit{D\textsubscript{2h}}) to monoclinic (\textit{C\textsubscript{2h}}) symmetry \cite{kaneko2013orthorhombic}. Within this framework, the charge-transfer excitons emerge from electron-hole pairing between Ta $5d_{xy}$ orbitals and hybridized Ni $3d_{xz+yz}$/Se $4p_{x+y}$ states within a single layer.  

Complementary x-ray diffraction (XRD) and Raman spectroscopy measurements confirm the high crystalline quality and phase purity of the samples (SM \cite{supplemental}, Sec. S1). The room-temperature Raman spectrum (SM \cite{supplemental} Fig.S3) shows multiple Raman-active phonon modes with frequencies and symmetries consistent with previously reported A\textsubscript{g} and B\textsubscript{2g} modes of TNSe \cite{yan2019strong,kim2020phononic,ye2021lattice}.

\subsection{\label{sec:levelB}Experimental setup}
We have employed non-degenerate pump-probe spectroscopy in reflection geometry to investigate the ultrafast dynamics of the sample (schematic shown in SM \cite{supplemental} Fig.S4). The differential reflectivity measurements are carried out using a regenerative amplifier system (RegA9050, Coherent Inc., USA) delivering pulses of $\sim60$~fs duration operating at a 250~kHz repetition rate with a central photon energy of $\sim1.57$~eV ($\sim795$~nm). The higher-intensity part of the fundamental beam is frequency-doubled in a type-II $\beta$-barium borate (BBO) crystal to generate the 3.14~eV pump, while the remaining 1.57~eV component is used as the probe. The transient differential probe reflectivity ($\Delta R/R$) is recorded using a balanced photodiode and lock-in amplifier. Throughout the experiments, the sample is mounted in a temperature-controlled closed chamber under a dry nitrogen atmosphere to minimize environmental effects. Temperature-dependent measurements are performed using a home-built heater, keeping pump and probe fluences fixed at $83$ and $7~\mu\mathrm{J/cm^2}$, respectively. Care has been taken to ensure that no measurable sample degradation occurred within our chosen temperature or fluence range. Additional experimental details are provided in the SM \cite{supplemental} Sec. S2. 

\section{\label{sec:level3}Results and Discussion}

\begin{figure*}
    \centering
    \includegraphics[width=\textwidth]{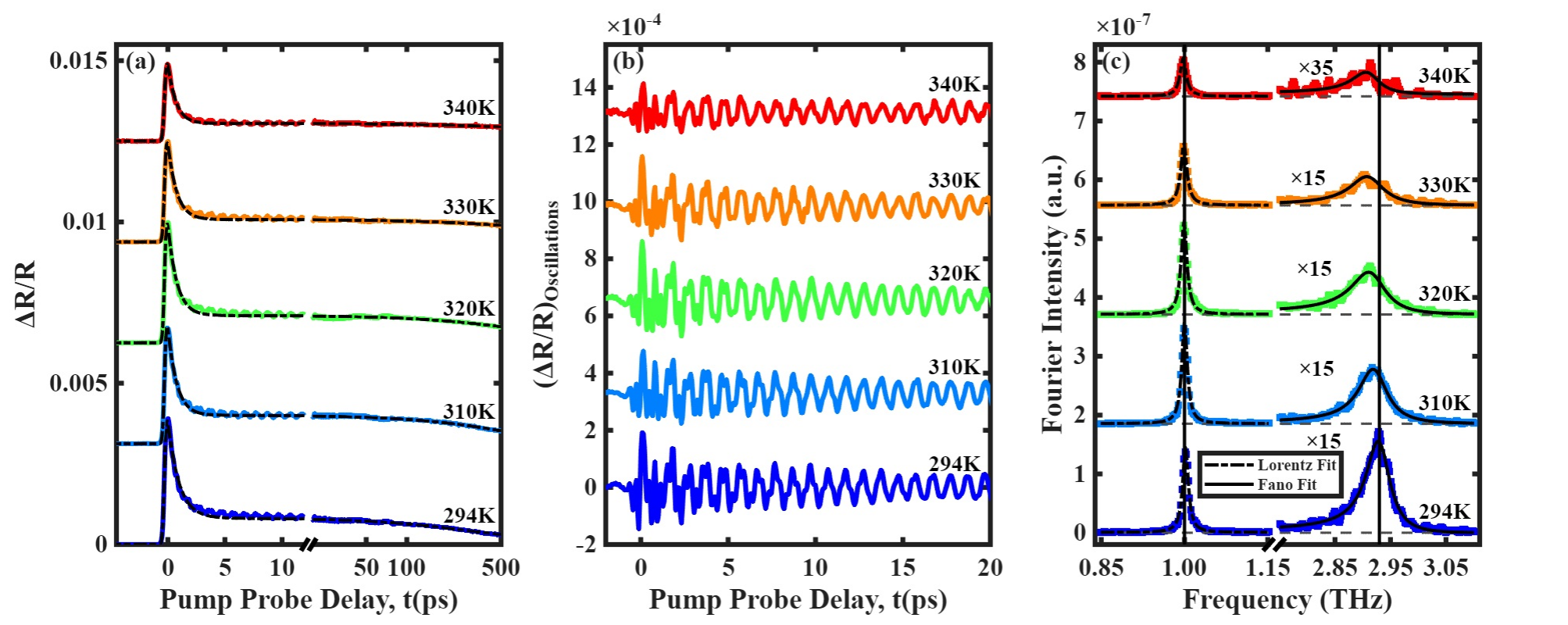}
    \begin{subfigure}{0\linewidth}
        \phantomcaption
        \label{fig:2a}
    \end{subfigure}
    \begin{subfigure}{0\linewidth}
        \phantomcaption
        \label{fig:2b}
    \end{subfigure}
    \begin{subfigure}{0\linewidth}
        \phantomcaption
        \label{fig:2c}
    \end{subfigure}
    \caption{\justifying (a) Temperature-dependent time-resolved differential reflectivity ($\Delta R/R$) signals of Ta\textsubscript{2}NiSe\textsubscript{5} measured at selected temperatures both below and above the transition temperature, \(T_c \approx 327\)~K. Colored solid lines represent experimental data at different temperatures, while black dash-dotted lines denote fits to the data using Eq.\eqref{eq:electronicbgfit}. The time axis is shown on a linear scale up to 15~ps and on a logarithmic scale at longer delays to highlight the full relaxation dynamics. (b) Coherent oscillations extracted by subtracting the fitted electronic background from the data in (a). (c) Fourier intensity (FFTs) of the oscillations, revealing dominant phonon modes at 1.0 and 2.9~THz (colored solid lines). For clarity, the Fourier intensities of the higher-frequency mode are scaled by factors of $\times15$ and $\times35$, as indicated. The FFT spectra are fitted using Lorentzian and Fano line shape functions, as shown by the black dash-dotted and solid lines, respectively. All the measurements were performed with a pump and probe fluence of $83$ and $7~\mu\mathrm{J/cm^2}$, respectively. Traces are vertically offset for better visualization.}
    \label{fig:2}
\end{figure*}

\subsection{\label{sec:levelA}Temperature dependence of carrier relaxation and coherent optical phonon dynamics}

The transient reflectivity data ($\Delta R/R$) up to a delay of 500~ps at selected temperatures are shown in Fig.\ref{fig:2a} (a complete set of temperature-dependent data is provided in SM \cite{supplemental} Sec. S3 Fig.S5). The measured time traces consist of a superposition of an electronic response and periodic oscillations arising from multiple coherent optical phonon modes. Here, the pump pulse excites electrons from the valence band to the higher-lying states in the conduction band, creating a population of hot carriers. The 1.57~eV probe, although well above the fundamental band gap, remains sensitive to the pump-induced modifications of the dielectric function through carrier-induced band renormalization and phonon-driven modulations, allowing us to track the resulting changes in reflectivity. The sharp positive peak observed at zero time delay between the pump and the probe (t = 0~ps) corresponds to the instantaneous generation of high-energy carriers. Following the initial positive peak, the signal exhibits a multi-exponential decay that captures the various relaxation channels across the band gap \cite{kabanov1999quasiparticle,werdehausen2018photo}. In order to quantify the photo-carrier dynamics and to isolate the coherent oscillatory components, we fit the electronic background using the following function:
    
\begin{equation}
    \frac{\Delta R(t)}{R} = \left[1+\textit{erf}\left(\frac{t-t_0}{\tau_{r}}\right)\right]\times\left[A_0 + A_1\exp\left(-\frac{t}{\tau_1}\right) + A_2\exp\left(-\frac{t}{\tau_2}\right)\right]
    \label{eq:electronicbgfit}
\end{equation}
where A\textsubscript{1} (A\textsubscript{2}) and $\tau_1$ ($\tau_2$) represent the amplitude and relaxation time constant associated with the fast (slow) dynamical process, respectively (refer to Fig.\ref{fig:3} in the next section for a more detailed discussion). The constant term, A\textsubscript{0}, represents a long-lived offset commonly attributed to effective bolometric contributions arising from laser-induced heat dissipation of the irradiated region. The prefactor, \begin{math}\left[1+\textit{erf}\left(\frac{t-t_0}{\tau_{r}}\right)\right]\end{math} in Eq.\eqref{eq:electronicbgfit}, accounts for the finite rise of the signal following photoexcitation. The extracted rise time, $\tau_r\approx191\pm2$~fs remains almost constant across all the temperatures [see SM \cite{supplemental} Fig.S6(b)] and is comparable to our instrument’s cross-correlation width of $\sim124$~fs. Hence, it is challenging to attribute this rise time to any underlying physical process within the sample, as it is likely limited by our experimental time resolution rather than reflecting an intrinsic ultrafast dynamic. Fig.\ref{fig:2b} presents the coherent optical phonon oscillations obtained by subtracting the fitted biexponential background (Eq.\eqref{eq:electronicbgfit}, black dash-dotted lines) from the raw $\Delta R/R$ signal (colored solid lines). These coherent oscillations are further analyzed in the frequency domain using fast Fourier transforms (FFTs), as shown in Fig.\ref{fig:2c}. Two prominent phonon modes are observed at 1.0 and 2.9~THz. To fit these modes, we have used Lorentzian and Fano line shapes, respectively, capturing the sharp symmetric response of the fundamental mode and the asymmetric line shape of the higher-frequency feature, indicative of coupling to an underlying continuum. This frequency-domain analysis provides a clear identification of the coherent optical phonon modes and their temperature-dependent behavior. In the following section, we analyze the temperature dependence of the excitonic relaxation dynamics and the associated coherent phonon responses in detail.

\subsubsection{\label{sec:leveli}Exciton relaxation dynamics}

\begin{figure*}
    \centering
    \includegraphics[width=\textwidth]{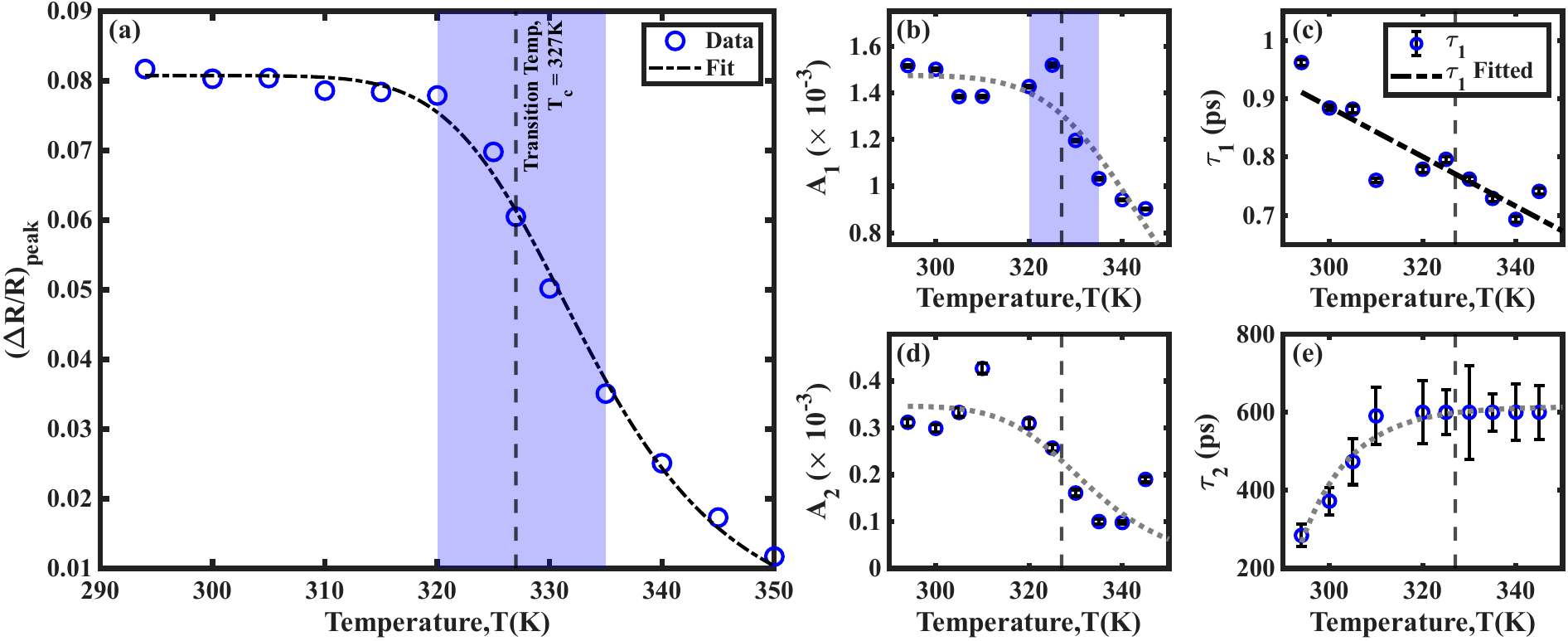}
    \begin{subfigure}{0\linewidth}
        \phantomcaption
        \label{fig:3a}
    \end{subfigure}
    \begin{subfigure}{0\linewidth}
        \phantomcaption
        \label{fig:3b}
    \end{subfigure}
    \begin{subfigure}{0\linewidth}
        \phantomcaption
        \label{fig:3c}
    \end{subfigure}
    \begin{subfigure}{0\linewidth}
        \phantomcaption
        \label{fig:3d}
    \end{subfigure}
    \begin{subfigure}{0\linewidth}
        \phantomcaption
        \label{fig:3e}
    \end{subfigure}
    \caption{\justifying (a) Temperature dependence of the peak transient reflectivity amplitude, $(\Delta R/R)_{peak}$. The signal remains nearly constant at low temperatures and decreases gradually as the system approaches the transition. The black dash-dotted line shows a fit using Eq.\eqref{eq:peakfit}. (b,d) Temperature dependence of the amplitudes, A\textsubscript{1} and A\textsubscript{2} and (c,e) the corresponding relaxation times $\tau_1$ and $\tau_2$ of the fast and slow relaxation processes, respectively, as defined in Eq.\eqref{eq:electronicbgfit}. For A\textsubscript{1}, the RT functional form [Eq.\eqref{eq:RTmodel}] is included only as a guide to the eye (gray dashed curve). The relaxation time $\tau_1$ decreases almost linearly with temperature, as indicated by the black dash-dotted line. No specific model is applied to A\textsubscript{2} and $\tau_2$; the gray dotted lines in (b), (d), and (e) are guides to the eye. The blue shaded region and black dashed vertical line highlight the transition across ($T_c \approx 327$)~K, and the data points are shown as blue circles.}
    \label{fig:3}
\end{figure*}

To gain quantitative insight into the carrier relaxation dynamics described above, we first analyze the temperature dependence of the peak transient reflectivity amplitude, $(\Delta R/R)_{peak}$, which represents the maximum photoinduced change in the carrier population immediately after excitation. As shown in Fig.\ref{fig:3a}, $(\Delta R/R)_{peak}$ remains nearly constant at low temperatures but starts decreasing gradually as the system approaches the transition temperature \(T_c \approx 327\)~K. This overall trend resembles the behavior predicted by the Rothwarf-Taylor (RT) model \cite{kabanov1999quasiparticle,demsar1999superconducting,mihailovic1999time,demser1999single,kabanov2005kinetics}, which relates the temperature dependence of the photoinduced reflectivity to the underlying energy gap associated with the ordered phase \cite{rothwarf1967measurement} (see SM \cite{supplemental} Sec. S4 for model details). Here, we employ a modified version of this model to account for the smoother suppression of $(\Delta R/R)_{peak}$ observed in our data,

\begin{equation}
    \left(\frac{\Delta R}{R}\right)_\mathrm{peak} = \frac{A/\mathcal{E}_\mathrm{g}}{1+B \exp\!\left(-\frac{\mathcal{E}_\mathrm{g}}{2k_\mathrm{B}\left(T-T_\mathrm{d}\right)}\right)} 
    \label{eq:peakfit}
\end{equation}

where A, B, and T\textsubscript{d} are free fitting parameters. The fit to Eq.\eqref{eq:peakfit} yields an effective excitonic-insulator energy scale of $\mathcal{E}_\mathrm{g} = 100\pm69$~meV, with an offset temperature of $T_d=272\pm22$~K. Within the uncertainty, the value of $\mathcal{E}_\mathrm{g}$ is consistent with an excitonic gap, $\Delta_\mathrm{EI} \equiv 2\Delta$ = 200~meV, which lies within the range ($\sim200-320$~meV) reported previously for TNSe \cite{lu2017zero,werdehausen2018photo}. While the uncertainty limits a precise determination of the gap magnitude, the model captures the overall temperature-dependent suppression of the peak amplitude, reflecting the gradual weakening of excitonic correlations with increasing temperature. This behavior is consistent with the energy scale inferred from transport measurements \cite{lu2017zero} and ultrafast amplitude dynamics in bulk TNSe \cite{werdehausen2018photo}. The modified RT model thus effectively provides a phenomenological description of how excitonic correlations evolve as the system approaches the transition, signaling the gradual melting of the EI phase.

We note, however, that the temperature dependence of the peak transient amplitude may also contain a contribution from the equilibrium optical response at the probe photon energy. The 1.57~eV probe lies close to interband transitions involving the $\Gamma$-point electronic structure, whose energies and optical matrix elements evolve across the phase transition. Indeed, the equilibrium reflectivity measured at 1.57~eV exhibits a temperature dependence similar to that of the peak $\Delta R/R$ (see SM \cite{supplemental}, Fig. S14). Consequently, although the RT-based analysis of the transient amplitude is consistent with the evolution of the correlation gap, its temperature dependence cannot be attributed exclusively to the photoexcited quasiparticle population; variations in the optical sensitivity of the 1.57~eV probe associated with the evolving interband transitions are also expected to contribute.

Figures \ref{fig:3b}-\ref{fig:3e} summarize the temperature dependence of the relaxation amplitudes A\textsubscript{1} and A\textsubscript{2}, along with the corresponding timescales $\tau_1$ and $\tau_2$ associated with the fast and slow relaxation channels, respectively. The fast relaxation time $\tau_1$ decreases linearly with increasing temperature, from $0.96\pm0.01$~ps at 294~K to $0.69\pm0.01$~ps at 340~K. This monotonic decrease contrasts with the behavior expected from the conventional RT model, where both the relaxation time and amplitude typically exhibit correlated temperature dependencies \cite{werdehausen2018photo,arora2025ultrafast}. The observed deviation suggests the presence of additional temperature-dependent scattering pathways beyond simple quasiparticle recombination across a static gap. Physically, $\tau_1$ reflects the timescale of the initial carrier relaxation following photoexcitation. The pump pulse transiently dissociates bound excitons, generating a non equilibrium distribution of unbound electron-hole pairs. Subsequently, $\tau_1$ captures the carrier cooling and recombination of these carriers, leading to the reformation of excitons and restoration of the excitonic condensate \cite{werdehausen2018photo,mor2022ultrafast}. Above $T_c$, $\tau_1$ is associated with the formation and relaxation of thermalized excitons in the semiconducting phase.
The amplitude A\textsubscript{1} quantifies the fraction of photoexcited carriers relaxing through this fast channel and thus serves as a measure of the condensate response in TNSe. Below $T_c$, A\textsubscript{1} remains nearly constant, indicating a robust and temperature-independent excitonic gap weakly perturbed by the pump. As the system approaches $T_c$, A\textsubscript{1} shows a pronounced drop, signaling the collapse of the gap, a behavior consistent with the mean-field-like order parameter. This trend is consistent with the behavior expected from the conventional RT model, which captures the suppression of the photoinduced response in gapped systems \cite{werdehausen2018photo,arora2025ultrafast}. To illustrate this, we overlay the RT functional form as a guide to the eye in Fig.\ref{fig:3b}.

\begin{equation}
    A_1\left(T\right) \propto \frac{1/\Delta}{1+D \exp\!\left(-\frac{\Delta}{2k_\mathrm{B}T}\right)}
    \label{eq:RTmodel}
\end{equation}

where $D$ is a phenomenological constant. A quantitative fit is not performed because the limited number of temperature points does not allow reliable constraints of the multi-parameter model. The curve, therefore, serves only as a visual reference, indicating that the observed drop in A\textsubscript{1} near $T_c$ is compatible with RT-like dynamics without assigning a precise gap value.

As shown in Fig.\ref{fig:3e}, the slow relaxation time $\tau_2$ increases from $284\pm28$~ps at 294 K to about $591\pm74$~ps near 310~K and then saturates at $\sim 600$~ps at higher temperatures. The corresponding amplitude A\textsubscript{2} grows gradually up to $\sim 305$~K, shows weak fluctuations near 310~K, and decreases thereafter. These trends indicate that the slow relaxation pathway becomes increasingly important when approaching the transition, but its relative contribution weakens again at higher temperatures.

In the present experiment, photoexcitation at 3.14~eV generates carriers far above the excitonic gap. These carriers rapidly thermalize through carrier-carrier interactions and subsequent energy transfer to the lattice on sub-picosecond timescales. As a result, the pump pulse primarily creates a non-equilibrium distribution of lattice excitations rather than directly dictating the long-time electronic recovery. The long relaxation time $\tau_2$, therefore, does not correspond to simple carrier recombination; instead, it reflects the delayed re-establishment of excitonic correlations in the presence of these slowly relaxing lattice degrees of freedom. In this sense, photoexcited carriers act mainly as the initial driver of lattice non-equilibrium, while the recovery bottleneck is governed by phonon-assisted exciton relaxation rather than direct carrier recombination.

\begin{table*}
    \centering
    \caption{Summary of reported slow relaxation times ($\tau_2$) from previous pump-probe studies on TNSe.}
    \begin{tabular}{>{\centering\arraybackslash}p{0.15\linewidth}>{\centering\arraybackslash}p{0.15\linewidth}>{\centering\arraybackslash}p{0.15\linewidth}>{\centering\arraybackslash}p{0.15\linewidth}p{0.3\linewidth}}\toprule
        \textbf{References} &  \textbf{Pump Wavelength} &  \textbf{Fitting Window} & \textbf{$\tau_2$ (ps)} & \textbf{Reported Origin}\\\midrule
        Ref. \cite{werdehausen2018photo} & 800~nm & 15~ps & $\sim6-50$ & Exciton breaking and carrier recombination\\
         Ref. \cite{liu2021photoinduced} & 800~nm & 12~ps & $\sim10$ & Phonon-phonon/dephasing-related relaxation\\
         Ref. \cite{miyamoto2022charge} & 800~nm & 5~ps & $\sim12$ & - \\
         Ref. \cite{jiang2024multiple} & 800~nm & 15~ps & $\sim2-5$ & -\\
         Ref. \cite{arora2025ultrafast} & 800~nm & 120~ps & $\sim2-12$ & Cooling of hot high-frequency phonons to low-frequency modes\\
         \textbf{Present Work} & \textbf{400~nm} & \textbf{500~ps} & \textbf{$\sim200-600$} & \textbf{Phonon bottleneck-dominated relaxation}\\ \bottomrule\end{tabular}
    \label{tab:tau2details}
\end{table*}

The monotonic increase and eventual saturation of $\tau_2$ indicate that the slow relaxation pathway becomes increasingly important as the system approaches the transition temperature. Notably, the extracted $\tau_2$ values, reaching several hundred picoseconds, are much longer than those reported previously ($\sim 2-50$~ps; see Table\ref{tab:tau2details}). This difference arises primarily from the substantially longer temporal window employed here (up to 500~ps), which allows us to resolve slow relaxation components that remain hidden in measurements restricted to 5-120~ps. By capturing the full decay of this channel, the extended delay range reveals a process that governs the late-time dynamics but would otherwise be truncated in shorter measurements (see SM \cite{supplemental} Sec. S3.4).

The temperature evolution of $\tau_2$ is consistent with a phonon-assisted relaxation bottleneck. A phenomenological description based on an exciton-phonon bottleneck model, analogous to the Rothwarf-Taylor framework, is provided in the SM \cite{supplemental} (Sec. S4). Such an interpretation is motivated by the strong coupling between excitonic and lattice degrees of freedom previously reported in TNSe. In particular, Larkin et al. \cite{larkin2017giant} interpreted the giant excitonic Fano resonance in terms of spatially extended exciton-phonon bound states, while subsequent theoretical work \cite{jiang2024multiple} has further investigated the coupling between excitonic and lattice degrees of freedom. These studies provide a physical basis for considering phonon-dressed or exciton-polaron-like excitations in TNSe. Within this phenomenological picture, the long-lived component reflects the recovery of the coupled excitonic-lattice system in the presence of a nonequilibrium phonon population. We emphasize that the present measurements do not establish a microscopic exciton-polaron mechanism; rather, the strong exciton-phonon coupling reported previously provides a physical motivation for the bottleneck interpretation adopted here. As temperature increases, thermally populated phonons promote reabsorption and reduce the net rate of energy dissipation, leading to progressively slower recovery as T approaches $T_c$. Similar bottleneck effects are well documented in other material systems, including hot-phonon-limited cooling in lead-halide perovskites \cite{yang2016observation,yang2017acoustic,faber2024hot}. In Ta$_2$NiSe$_5$, however, the relevant degrees of freedom are excitons rather than free carriers, so $\tau_2$ reflects the phonon-mediated recovery of excitonic correlations. At lower temperatures relaxation proceeds more rapidly, whereas near and above $T_c$, the weakening of excitonic correlations results in saturation of $\tau_2$. The amplitude A\textsubscript{2} reflects the relative weight of this slow pathway; although its temperature trend is broadly consistent with enhanced phonon involvement, we do not assign it a quantitative microscopic interpretation. Taken together, these observations indicate that long-lived phonon-assisted relaxation channels, often inaccessible in shorter time-window measurements, play an important role in the non-equilibrium dynamics of Ta$_2$NiSe$_5$ near the transition. 

In our measurements, the parameter A\textsubscript{0} appears as a constant, long-lived offset in the transient reflectivity signal [(]see SM \cite{supplemental} Fig.S6(a)]. This offset is attributed to a bolometric contribution arising from lattice heating in the photoexcited region. The deposited energy diffuses gradually through the material via bulk thermal diffusion, which describes macroscopic heat flow through the lattice rather than microscopic carrier or phonon relaxation. The corresponding thermal diffusion time can be estimated from $t_{diff} = L^2/\alpha$; where $L\approx$20-25~nm is the optical penetration depth of the 400~nm pump and $\alpha\sim1.3\times10^{-5}$~m\textsuperscript{2}/s is the thermal diffusivity (see SM \cite{supplemental} Sec. S3.3 for calculations). This yields t\textsubscript{diff} = 30-50~ps, comparable in magnitude to the relaxation time $\tau_2$ but fundamentally distinct in origin. While vertical heat redistribution may begin on sub-nanosecond timescales, lateral diffusion across the laser spot ($\sim60~\mu$m) would occur on the millisecond scale. Consequently, the bolometric background remains effectively time-independent within our 0-500 ps measurement window. Similar long-lived offsets have been reported in ultrafast pump-probe studies of TNSe \cite{werdehausen2018photo,arora2025ultrafast}.

Overall, the excitonic relaxation dynamics revealed here consist of an initial ultrafast recovery of excitonic correlations followed by a long-lived, temperature-dependent exciton-phonon bottleneck. These findings are consistent with the photodoping behavior observed in tr-ARPES studies of TNSe \cite{mor2017ultrafast} and highlight the importance of extended temporal windows for capturing slow relaxation processes near the excitonic-insulator transition.

\subsubsection{\label{sec:levelii}Coherent phonon dynamics}
\begin{figure*}
    \centering
    \includegraphics[width=\textwidth]{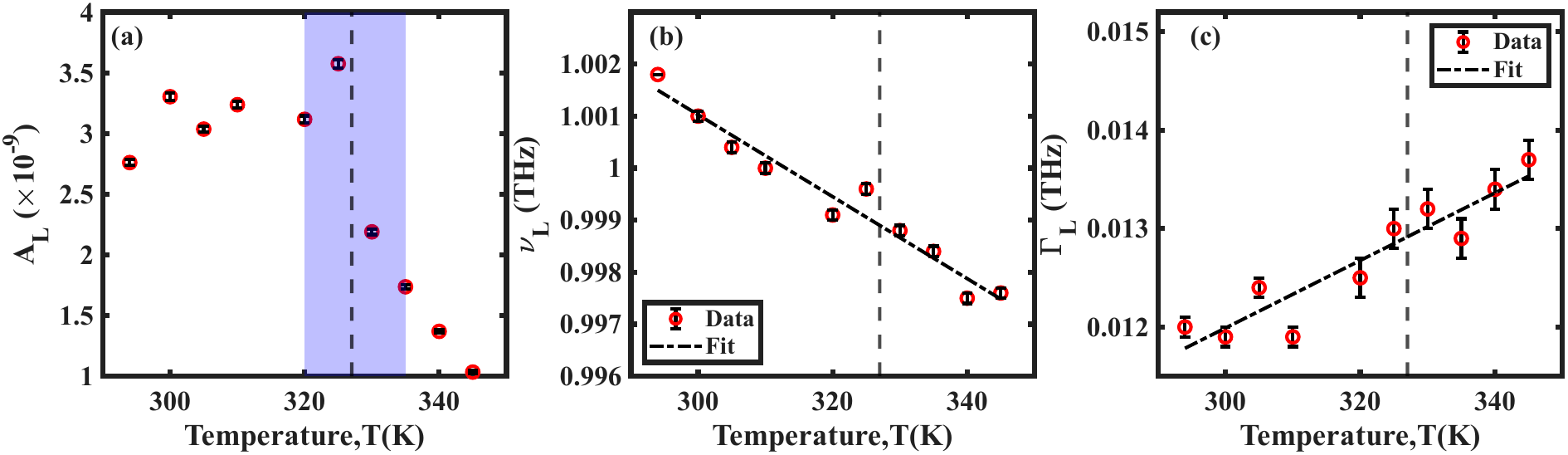} 
    \begin{subfigure}{0\linewidth}
        \phantomcaption
        \label{fig:4a}
    \end{subfigure}
    \begin{subfigure}{0\linewidth}
        \phantomcaption
        \label{fig:4b}
    \end{subfigure}
    \begin{subfigure}{0\linewidth}
        \phantomcaption
        \label{fig:4c}
    \end{subfigure}
    
    \caption{\justifying (a) Temperature dependence of the amplitude of the 1.0 THz mode (CP1), A\textsubscript{L}, which is strongly coupled to the excitonic condensate. (b) and (c) Temperature dependence of the frequency ($\nu_L$) and linewidth ($\Gamma_L$) of the CP1 mode, respectively. The dash-dotted lines are fits to an anharmonic decay model. The shaded region highlights the vicinity of the phase transition, and the vertical dashed line marks the critical temperature $T_c$.}
    \label{fig:4}
\end{figure*}

The electronic relaxation processes discussed above are intrinsically coupled to the lattice through strong exciton-phonon interactions \cite{yan2019strong,werdehausen2018coherent}. Consequently, photoexcitation perturbs not only the excitonic condensate but also drives coherent optical phonons that serve as fingerprints of the coupled electronic and structural dynamics. The observed modes at approximately 1.0~THz (CP1), 2.9~THz (CP2) and 3.7~THz (CP3) are consistent with previously reported Raman-active vibrations in TNSe \cite{yan2019strong,kim2021direct,ye2021lattice} and with modes identified in earlier ultrafast optical studies \cite{werdehausen2018coherent,miyamoto2022charge,jiang2024multiple,wang2026pressure}. The CP3 mode exhibits a very weak amplitude and is therefore not shown in Fig.\ref{fig:2c}, but it is presented in SM \cite{supplemental} Fig.S5(c). In the following, we focus on the temperature evolution of the dominant CP1 and CP2 modes, which provides quantitative insight into the evolution of exciton-phonon coupling across the excitonic insulator regime.

Among these modes, CP1 is the most prominent and has been widely associated with the collective amplitude oscillation of the excitonic order parameter, a Higgs-like mode that modulates the Ni-Se bond length along the chain direction \cite{werdehausen2018coherent,chu2022anisotropic} [see SM \cite{supplemental}, Fig.S9(a)]. The resulting coherent oscillation amplitude of CP1 reflects the dynamical modulation of the excitonic order parameter and is expected to weaken as the temperature approaches $T_c$. In contrast, the CP2 and CP3 modes arise from conventional optical phonons: The former involves Se displacements perpendicular to the Ta-Ni-Se layers, while the latter reflects mixed Ta-Se vibrations within the layers \cite{nakano2018antiferroelectric,ye2021lattice,baldini2023spontaneous} (see SM \cite{supplemental} Sec. S3.6).

In the frequency domain, the CP1 mode exhibits a symmetric spectral profile that is well captured by a Lorentzian function given by Eq.\eqref{eq:Lorentz}. In contrast, the CP2 mode displays a weak asymmetric Fano-like broadening [see Eq. \eqref{eq:Fano}] that indicates the quantum interference between the discrete phonon and a transient electronic continuum generated by photoexcited carriers \cite{fano1961effects,hase2006photoinduced,kumar2007photoexcited}.
While the asymmetry is subtle (see SM \cite{supplemental} Sec. S3.5 for detailed analysis), the extracted asymmetry parameter remains finite across the measured temperature range, without a clear systematic temperature dependence within experimental uncertainty. To characterize the temperature evolution, the FFT spectra of CP1 and CP2 are fitted using Lorentzian and Breit-Wigner-Fano (BWF) lineshapes, respectively. The corresponding fits are shown in Fig.\ref{fig:2c} as dash-dotted (Lorentzian) and solid black (BWF) curves. Prior to fitting, FFT band-pass filters were applied around each phonon frequency to isolate individual modes. This bandwidth was chosen to fully encompass the phonon linewidths while suppressing adjacent modes and background noise, ensuring that the intrinsic peak shape remains unaffected. A more detailed comparison of Lorentzian and Fano fits for the CP2 mode across the selected temperatures, including the extracted frequencies, linewidths, and fit quality ($R_s^2$), is provided in the SM \cite{supplemental} (Fig.S8 and Table TS3).
\begin{equation}
    I_L(\omega) = \frac{2A_L/\pi \Gamma_L}{1+\left[\left(\omega-\omega_L\right)/\left(\Gamma_L/2\right)\right]^2}
    \label{eq:Lorentz}
\end{equation}

\begin{equation}
    I_F(\omega) = A_F\frac{\left[1+\left(\omega-\omega_F\right)/\left(q\Gamma_F\right)\right]^2}{1+\left[\left(\omega-\omega_F\right)/\Gamma_F\right]^2}
    \label{eq:Fano}
\end{equation}

Here, $A_{L(F)},\Gamma_{L(F)}$, and $\omega_{L(F)}$ are the amplitude, linewidth, and peak frequency of the Lorentz (Fano) lineshape function. The asymmetry parameter |1/q| in Eq.\eqref{eq:Fano} quantifies the coupling strength between the phonons and the photoexcited electronic continuum. 

\begin{figure}
    \centering
    \includegraphics[width=0.8\textwidth]{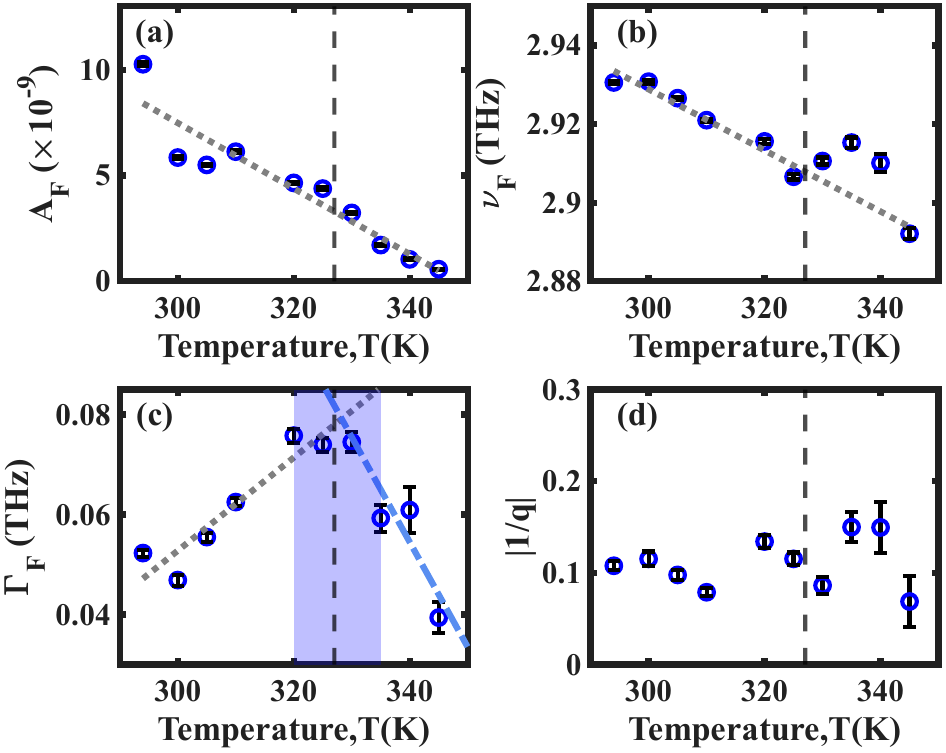} 
    \begin{subfigure}{0\linewidth}
        \phantomcaption
        \label{fig:5a}
    \end{subfigure}
    \begin{subfigure}{0\linewidth}
        \phantomcaption
        \label{fig:5b}
    \end{subfigure}
    \begin{subfigure}{0\linewidth}
        \phantomcaption
        \label{fig:5c}
    \end{subfigure}
    \begin{subfigure}{0\linewidth}
        \phantomcaption
        \label{fig:5d}
    \end{subfigure}
    
    \caption{\justifying (a) Temperature dependence of the FFT amplitude of the 2.9~THz phonon mode (CP2),  A\textsubscript{F}, which does not couple directly to the excitonic order parameter (gray dotted line is a guide to the eye). (b) and (c) Temperature dependence of the  frequency ($\nu_F$) and linewidth ($\Gamma_F$) of the CP2 mode. Dash-dotted curves denote anharmonic decay fits for $T<T_c$. Extrapolation beyond $T_c$ reveals systematic deviations from the anharmonic trend, including anomalous linewidth narrowing in the high-temperature phase, indicative of a structural phase transition. In (c), a linear fit to the data for $T>T_c$ is shown as a light blue dash-dotted line. (d) Absolute value of the Fano parameter, $\left|1/q\right|$, as a function of temperature, quantifying the coupling between the phonon and the electronic continuum. The shaded region highlights the vicinity of the phase transition, and the vertical dashed line marks the critical temperature $T_c$.}
    \label{fig:4}
\end{figure}

 The temperature evolution of these parameters is summarized in Figs.\ref{fig:4a}-\ref{fig:4c} and Figs.\ref{fig:5a}-\ref{fig:5d}. The amplitude of the CP1 mode (A\textsubscript{L}) remains nearly constant over the range 294-327~K and decreases systematically at higher temperatures [Fig.\ref{fig:4a}]. This behavior qualitatively resembles an order-parameter-like suppression near the excitonic-insulator transition but does not exhibit the sharp critical collapse expected for a mean-field condensate response. Structurally, the 1.0~THz vibration corresponds to the in-phase motion of the Ni and neighboring Ta chains, with adjacent Ta-Ni-Ta blocks oscillating out of phase. Within each block, the relative Ni-Ta distance remains nearly unchanged, while the Se atoms shift significantly relative to the metal sites. Such Ni-Se displacements modulate local bonding and the electronic structure, influencing the exciton binding energy and thereby coupling the phonon to the excitonic order parameter \cite{werdehausen2018coherent,jiang2024multiple}. The gradual reduction of the CP1 amplitude A\textsubscript{L} with increasing temperature is consistent with a progressive reduction in excitonic coherence. In contrast to high-fluence studies \cite{werdehausen2018coherent}, where strong photoexcitation drives substantial depletion of the excitonic condensate and reveals its nonlinear response, our measurements are performed at a relatively low pump fluence of $83~\mu$J/cm$^2$. A quantitative estimate of the pump-induced transient lattice heating (see Sec.~S3.3 of the SM \cite{supplemental}) gives a temperature rise of approximately 6~K at the employed fluence. The pump absorption depth is smaller than the probe sampling depth, resulting in a depth-averaged transient response, while the transition remains centered at $T_c\approx327$~K with no resolvable pump-induced shift [Fig.\ref{fig:3}(a)]. These observations indicate that the present measurements probe the system in a weakly perturbed regime without evidence for a significant photoinduced displacement of the phase boundary. This weakly perturbed regime allows us to investigate the recovery of the equilibrium excitonic state without deliberately driving the system into a strongly depleted state. The extended temporal window further enables us to examine the 1.0~THz coherent phonon response together with the long-lived electronic recovery observed in the same measurements.
 
 We therefore do not attempt a mean-field fit to A\textsubscript{L}. Instead, the observed suppression of A\textsubscript{L} with temperature is attributed to a thermally driven weakening of the excitonic condensate, which progressively reduces the coupling efficiency between the phonon and the electronic order parameter. The persistence of a finite oscillation amplitude below $T_c$ is consistent with the picture that, at low excitation densities, the condensate effectively remains confined in a deep potential well and cannot be driven fully into the nonlinear coupling regime \cite{werdehausen2018coherent}.

The frequency ($\nu_\mathrm{L}$) and linewidth ($\Gamma_\mathrm{L}$) of CP1, shown in Figs.\ref{fig:4b} and \ref{fig:4c}, respectively, exhibit a gradual softening and broadening with increasing temperature. Both trends follow the standard anharmonic decay behavior \cite{klemens1966anharmonic}, as confirmed by the following fit equations:
\begin{subequations}
\begin{align}
    \nu_{\text{cubic}}(T) & = \nu_0 - A\left[1+2n\left(\tfrac{\nu_0}{2}\right)\right]
\end{align}
\begin{align}
    \Gamma_{\text{cubic}}(T) & = \Gamma_0 + B\left[1+2n\left(\tfrac{\nu_0}{2}\right)\right].
\end{align}
\end{subequations}
where $\nu_0$ and $\Gamma_0$ are the zero-temperature values, $A$ and $B$ quantify the anharmonic coupling, and $n(\nu) = \frac{1}{\exp\left(\frac{h\nu}{k_BT}\right)-1}$ is the Bose-Einstein occupation number. These results are consistent with previous time-resolved optical \cite{werdehausen2018coherent, mor2018inhibition} and tr-ARPES \cite{seki2014excitonic} studies, which reported that the 1.0~THz mode amplitude tracks the excitonic order parameter and follows its mean-field-like temperature dependence.

The CP2 mode, however, behaves distinctly. The amplitude (A\textsubscript{F}), shown in Fig.\ref{fig:5a}, decreases nearly linearly with temperature, exhibiting no anomaly near $T_c$. The 2.9~THz vibration involves out-of-plane Se atom displacement perpendicular to the Ta-Ni-Se layers, which weakly perturbs the interchain overlap and thus does not couple strongly to the excitonic condensate. The monotonic reduction of A\textsubscript{F} is therefore attributed to enhanced lattice anharmonicity and phonon-phonon scattering with temperature.

The temperature evolution of the CP2 frequency ($\nu_F$) and linewidth ($\Gamma_F$) is presented in Figs.\ref{fig:5b} and \ref{fig:5c}, respectively. The frequency softens monotonically with increasing temperature and follows the expected anharmonic trend in the low-temperature phase ($T<T_c$). Accordingly, $\nu_F$ was fitted using a cubic anharmonic decay model below $T_c$. When extrapolated beyond $T_c$, however, the frequency shows systematic deviations from the anharmonic expectation.
A reliable anharmonic fit to the linewidth, $\Gamma_F$, could not be performed due to the limited number of data points available below $T_c$ (only six points), which restricts the robustness of the extracted parameters. Nevertheless, $\Gamma_F$ shows a pronounced anomaly near the transition. The linewidth broadens on approaching $T_c$, peaks close to the transition, and narrows again in the high-temperature phase. For $T>T_c$, $\Gamma_F$ exhibits an approximately linear trend [Fig.\ref{fig:5c}], as indicated by a linear fit shown as a guide to the eye. Notably, the linewidth increases by $\sim27$~GHz between 300 and 325~K, comparable to the Raman mode broadening ($\sim30$~GHz) reported by Yan \textit{et al.} \cite{yan2019strong}.
The nonmonotonic evolution of $\Gamma_F$ cannot be captured by a purely anharmonic phonon-phonon interaction model and points to an additional contribution associated with the structural phase transition. Together with the deviation of $\nu_F$ from anharmonic expectations above $T_c$, these results indicate a modification of the lattice potential and phonon self-energy across the transition, consistent with a structural component to the excitonic insulator phase. The corresponding fitting parameters are summarized in Table \ref{tab:anharmonicdecayfit}. The larger anharmonic coefficients obtained for CP2 compared to CP1 reflect the stronger sensitivity of this mode to lattice anharmonicity and structural fluctuations, particularly in the high-temperature phase.

\begin{table}[t]
    \centering
    \caption{\justifying Fitting parameters obtained from anharmonic decay model fits of phonon frequency and linewidth.}
    \begin{tabular}{|>{\centering\arraybackslash}p{0.18\linewidth}|>{\centering\arraybackslash}p{0.32\linewidth}|>{\centering\arraybackslash}p{0.32\linewidth}|} \hline
        & \textbf{CP1} &\textbf{CP2}\\ \hline
        \textbf{$\nu_0$ (THz)} & $1.025 \pm 0.002$  &$3.165 \pm 0.031$
\\ \hline
        \textbf{A (THz)} & $(0.969 \pm 0.076) \times 10^{-3}$  &$(2.968 \pm 0.412) \times 10^{-2}$
\\ \hline
        \textbf{$\Gamma_0$ (THz)} & $(0.168 \pm 0.152) \times 10^{-2}$  &-\\ \hline
        \textbf{B (THz)} & $(0.422 \pm 0.058) \times 10^{-3}$  &-\\ \hline
    \end{tabular}
    \label{tab:anharmonicdecayfit}
\end{table}

Finally, the temperature dependence of the Fano asymmetry parameter $\left|1/q\right|$ for the CP2 mode is shown in Fig.\ref{fig:5d}. Across most of the measured temperature range, the absolute value of the Fano asymmetry parameter $|1/q|$ remains small and only weakly temperature dependent, indicating that the CP2 mode is weakly coupled to an electronic continuum. A modest enhancement of $|1/q|$ is observed in the vicinity of the transition temperature $T_c$, consistent with increased electronic and lattice fluctuations near the phase transition. Importantly, the absence of any pronounced anomaly or critical behavior in $|1/q|$ indicates that phonon-continuum interference remains a secondary effect and does not play a dominant role in the relaxation dynamics. Overall, the small magnitude of $|1/q|$ support for the view that CP2 remains primarily lattice-dominated across the entire temperature range.

Overall, CP2 behaves as an essentially uncoupled coherent phonon whose dynamics are dictated by lattice anharmonicity, in contrast to CP1, which closely follows the excitonic order parameter.

Additional pump-fluence-dependent transient reflectivity measurements at temperatures 294 K (below $T_c$), 325 K (near $T_c$) and 340 K (above $T_c$) with pump fluence ranging from $21$ to $416~\mu\mathrm{J/cm^2}$ and at a fixed probe fluence of $7~\mu\mathrm{J/cm^2}$, indicate that carrier dynamics show visible changes with pump fluence whereas the phonon relaxation dynamics remain unaffected (see SM \cite{supplemental} Sec. S5 for the details). Specifically, $\tau_1$ increases marginally with pump fluence across all three temperatures, consistent with the findings in Ref.\cite{mor2022ultrafast}. Meanwhile, $\tau_2$ initially increases with fluence before saturating at higher excitation densities.

\section{\label{sec:level4}Conclusion}
We have investigated the non-equilibrium dynamics of Ta\textsubscript{2}NiSe\textsubscript{5} using temperature-dependent, non-degenerate pump-probe spectroscopy with an extended delay window of up to 500~ps. This approach resolves two distinct relaxation channels: a fast component ($\sim0.7-0.9$~ps) associated with carrier cooling and exciton reformation, which weakens as the system approaches $T_c$, and a much slower component with decay times of $\sim280-600$~ps, not captured in earlier studies, which were limited to shorter delay windows. This previously inaccessible timescale points to a slow recovery of excitonic correlations, consistent with an exciton-phonon bottleneck, in which reabsorption of nonequilibrium lattice excitations delays the restoration of equilibrium excitonic correlations.

In addition, we observe two coherent optical phonon modes at 1.0 and 2.9~THz with qualitatively different coupling mechanisms. The 1.0~THz mode is strongly coupled to the excitonic condensate and exhibits an order-parameter-like suppression near $T_c$, consistent with its role as a collective exciton-phonon mode. In contrast, the 2.9~THz mode does not couple directly to the excitonic order parameter; its frequency softening, linewidth evolution and finite Fano asymmetry near and above $T_c$ primarily reflect anharmonic lattice dynamics associated with the structural phase transition, with phonon-continuum interference playing only a secondary role.

Overall, our results demonstrate that long-lived phonon-mediated processes play a central role in the ultrafast dynamics of Ta\textsubscript{2}NiSe\textsubscript{5}. Although our measurements do not distinguish whether the equilibrium phase transition is driven primarily by an electronic or excitonic or a structural instability, they demonstrate that the excitonic and lattice degrees of freedom are strongly coupled in the nonequilibrium dynamics. The distinct behavior of the 1.0 and 2.9 THz modes, together with the long-lived exciton-phonon bottleneck, is consistent with an intertwined exciton-lattice picture rather than a purely electronic or purely structural description. Our results therefore highlight the importance of considering the coupled electronic and lattice dynamics when describing the phase transition and its subsequent recovery. This work thus provides a unified dynamical framework for understanding excitonic insulators beyond the short-time regime.

\section{\label{sec:level4}Acknowledgement}

The authors would like to thank the Ministry of Education (MoE), Government of India, for funding and IISER Kolkata for the infrastructural support to carry out the research. P.G. would like to acknowledge the Department of Science and Technology (DST)-Innovation in Science Pursuit for Inspired Research (INSPIRE) for support from the research fellowship. A.C. acknowledges the Council of Scientific and Industrial Research (CSIR); S.S. acknowledges the University Grants Commission (UGC); S.K. acknowledges IACS for support from the Ph.D. fellowship. The authors also thank Pedisetti Venkatesh (former BS-MS student, IISER Kolkata) for assistance in the experimental work, including the setup of the heating sample stage. N.K. thanks ANRF, Govt. of India for funding through Research Grant ANRF/ARG/2025/004414/PS.

\section{\label{sec:level5}Data Availability}

The data that support the findings of this article are openly available \cite{data_availability}, embargo periods may apply.

\nocite{*}
\bibliography{References_Final}

\clearpage
\setcounter{figure}{0}
\setcounter{table}{0}
\setcounter{section}{0}
\setcounter{subsection}{0}
\setcounter{equation}{0}

\renewcommand{\thefigure}{S\arabic{figure}}
\renewcommand{\thetable}{TS\arabic{table}}
\renewcommand{\theequation}{S\arabic{equation}}
\renewcommand{\thesection}{S\arabic{section}}
\renewcommand{\thesubsection}{S\arabic{section}.\arabic{subsection}}
\renewcommand{\thesubsubsection}{S\arabic{section}.\arabic{subsection}.\arabic{subsubsection}}

\setcounter{secnumdepth}{2}
\setcounter{tocdepth}{2}

\section*{Supplemental Material}
\section{Sample Growth and Characterization}
\subsection{Sample Synthesis}
\begin{figure}[h]
  \centering
  \includegraphics[width=0.8\textwidth]{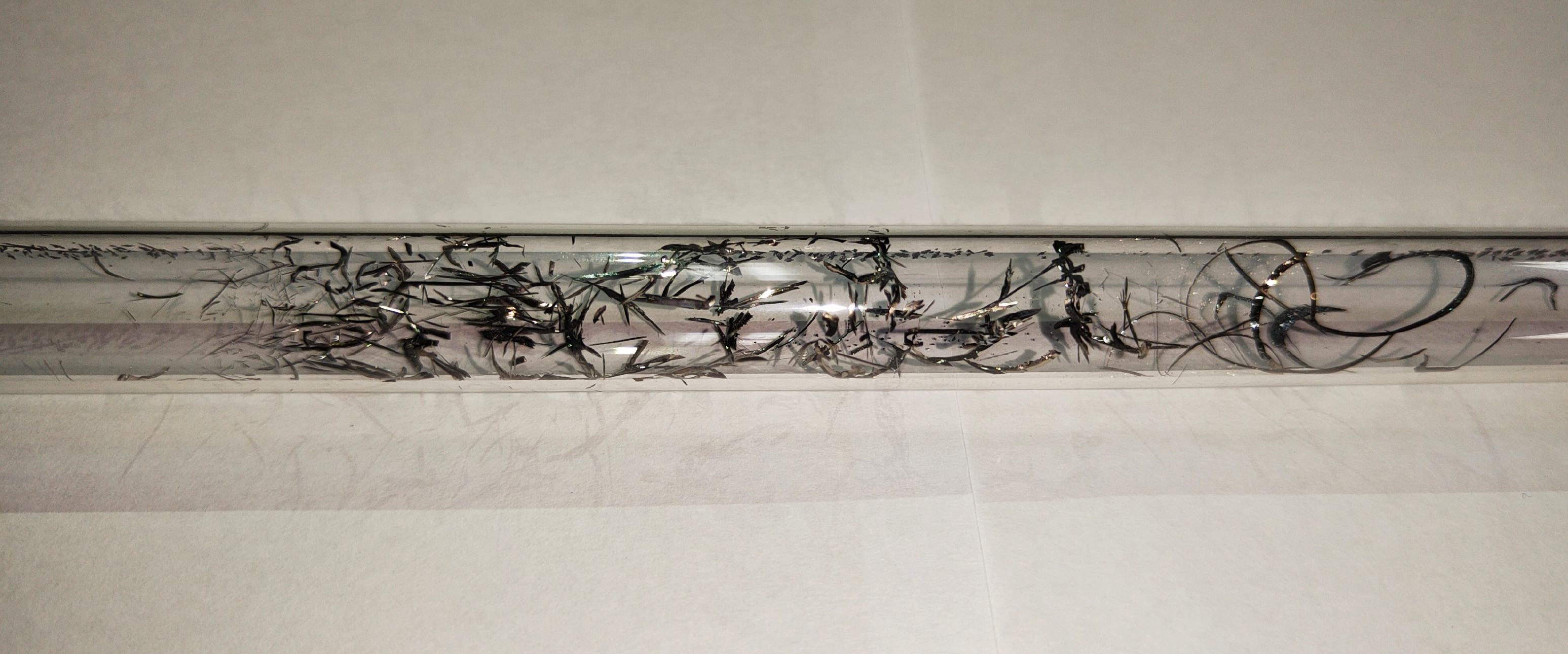}
  \caption{\centering Image of as-grown single crystals of Ta\textsubscript{2}NiSe\textsubscript{5}.}
  \label{fig:S1}
\end{figure}

Plate-like single crystals of Ta$_2$NiSe$_5$ (TNSe) were grown using the chemical vapor transport (CVT) method. High-purity elemental powders of tantalum, nickel, and selenium were combined in the appropriate stoichiometric ratios. The mixture was placed in a quartz ampoule, evacuated to approximately $1\times10^{-5}$~mbar, and sealed after the addition of a small quantity (3-4 $mg/cm^3$) of iodine $I_2$ as the transport agent. The sealed quartz tube was placed in a two-zone furance for 10 days under a temperature gradient of 950$^\circ$C and 850$^\circ$C with the powder in the source zone. Finally, the zones were cooled to room temperature, resulting in the formation of needle-like crystals in the cooler zone of the ampoule. The image of the as grown crystals inside the sealed ampoule is shown in Fig.\ref{fig:S1}.

\subsection{X-Ray Diffraction (XRD)}
\begin{figure}[h]
  \centering
  \includegraphics[width=0.8\textwidth]{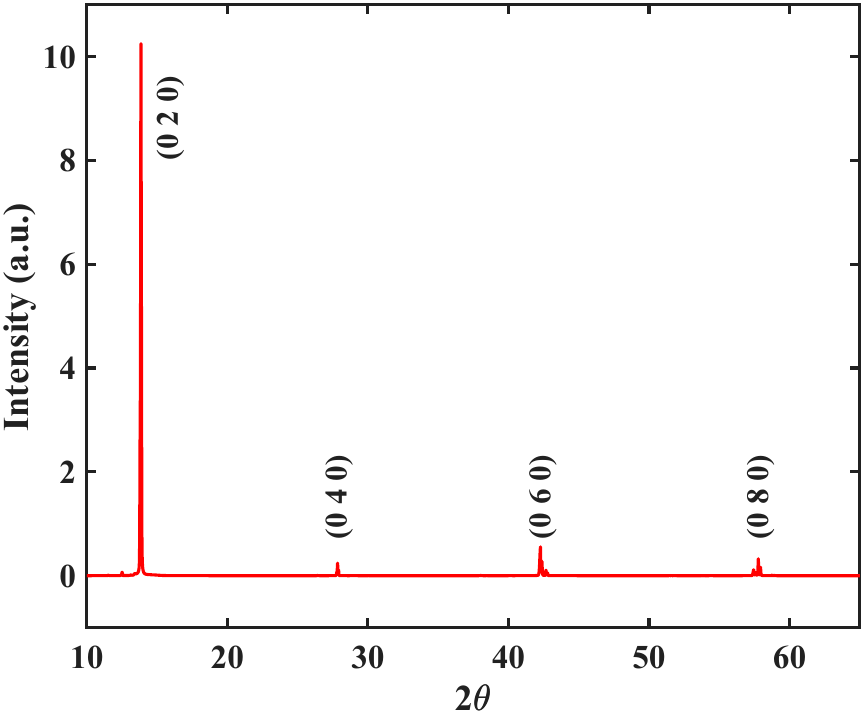}
  \caption{\justifying Room-temperature XRD data for the Cu K$_\alpha$ X-ray source projected on the ab plane of a single crystal of Ta\textsubscript{2}NiSe\textsubscript{5}.}
  \label{fig:S2}
\end{figure}

As shown in Fig.\ref{fig:S2}, XRD measurement was conducted to evaluate the phase purity and structure of the TNSe bulk crystal. The strongest diffraction peak at $2\theta=13.80^\circ$ can be matched with the (0 2 0) plane. In addition, other corresponding periodic diffraction peaks at $2\theta=27.85^\circ$, $42.32^\circ$ and $57.80^\circ$ can be well indexed to the (0 4 0), (0 6 0) and (0 8 0) planes, respectively. All diffraction peak positions can be matched with the monoclinic space group, C2/c.

\subsection{Temperature-Dependent Resistivity and Activation Energy Measurements}
To elucidate the transport behavior, we analyze the temperature-dependent resistivity in terms of an effective activation energy. For thermally activated transport (such as small polaron hopping), resistivity is commonly described by the Arrhenius relation \cite{laidler1984arrhenius}:
\begin{equation}
    \rho \sim \rho_0 \exp\left(\frac{E_\rho}{k_B T}\right)
\end{equation}
where, $E_\rho$ is the activation energy. In the simplest case, $E_\rho$ is assumed to be temperature independent and can be extracted from a linear fit of $\ln(\rho)$ versus (1/T). However, in correlated systems such as TNSe, where electronic structure and lattice degrees of freedom evolve with temperature, particularly across a phase transition, this assumption may not hold. To account for this, we evaluate the local activation energy \cite{seki2014excitonic,yan2019strong,lu2017zero,liu2021photoinduced}, defined as:
\begin{equation}
    E_\rho(T) = -k_BT^2 (\partial \ln(\rho)/\partial T)
\end{equation}
This approach enables a direct extraction of the activation energy from the experimental data without assuming it to be constant, allowing variations in the transport behavior to be tracked as a function of temperature. In addition, we determine the transition temperature ($T_C$) from the maxima of the $E_\rho$ vs $T$ plot. A constant $E_\rho$ corresponds to simple thermally activated behavior, whereas a temperature-dependent $E_\rho(T)$ indicates deviations arising from effects such as gap evolution, electron-lattice coupling, or disorder ~\cite{lu2017zero,di1986physical,seki2014excitonic,yan2019strong}.
As shown in Fig. 1(b) in the main manuscript, $E_\rho(T)$ exhibits a clear variation across the measured temperature range, with a distinct feature near \(T_c \approx 327\)~K, a behavior consistent with the anomaly observed in the resistivity data as discussed in the main manuscript.

\subsection{Raman Spectroscopy}
At the Brillouin zone center, the irreducible representation of TNSe yields (11A\textsubscript{g} + 11A\textsubscript{u} + 13B\textsubscript{g} + 13B\textsubscript{u}) \cite{yan2019strong}. Among the 45 optical phonon modes, 24 modes are Raman-active and the remaining 21 modes are infrared-active \cite{werdehausen2018coherent}.

\begin{figure}[h]
  \centering
  \includegraphics[width=0.8\textwidth]{S3.pdf}
  \caption{\justifying Room-temperature Raman spectrum (black line) of Ta\textsubscript{2}NiSe\textsubscript{5} single crystal, along with its multi-Lorentzian fit (red line). The observed peaks are labeled as M1-M8.}
  
  \label{fig:S3}
\end{figure}

Raman spectroscopy has  been carried out on bulk TNSe single crystals using a Horiba Jobin Yvon HR800 spectrometer with 532 nm laser excitation. The scattered light has been collected through a 50x objective and analyzed using a 600-grooves/mm diffraction grating. The resulting spectrum, shown in Fig.\ref{fig:S3}, is fitted with multiple Lorentzian functions, revealing eight distinct phonon modes at 60.37 (B\textsubscript{2g}), 94.45 (A\textsubscript{g}), 120.03 (A\textsubscript{g}), 173.01 (A\textsubscript{g}), 187.57 (A\textsubscript{g}), 210.51 (A\textsubscript{g}), 236.99 (A\textsubscript{g}) and 282.13 (A\textsubscript{g}) cm\textsuperscript{-1}, which are consistent with previously reported results \cite{yan2019strong,kim2020phononic,ye2021lattice}. The extracted peak frequencies and full width at half maximum (FWHM) values are summarized in Table \ref{tab:TS1}.

\begin{table}[h]
    \centering    \begin{tabular}{|c|c|c|c|}\hline
        \textbf{Mode} & \textbf{Wave number (cm\textsuperscript{–1})}& \textbf{FWHM (cm\textsuperscript{–1})} & \textbf{Symmetry Mode} \\
        \hline
        M1 & 60.4 $\pm$ 0.3& 13.1 $\pm$ 1.0& B\textsubscript{2g}  \\\hline
        M2 & 94.4 $\pm$ 0.0& 5.0 $\pm$ 0.0& A\textsubscript{g}  \\\hline
        M3 & 120.0 $\pm$ 0.0& 5.9 $\pm$ 0.1& A\textsubscript{g}  \\\hline
        M4& 173.0 $\pm$ 0.8& 10.0 $\pm$ 2.6& A\textsubscript{g}  \\\hline
        M5& 187.6 $\pm$ 0.6& 9.6 $\pm$ 2.4& A\textsubscript{g}  \\\hline
        M6& 210.5 $\pm$ 1.0& 14.1 $\pm$ 3.7& A\textsubscript{g}  \\ \hline
        M7& 237.0 $\pm$ 0.6& 20.0 $\pm$ 2.0&A\textsubscript{g}  \\\hline
        M8& 282.1 $\pm$ 0.4& 13.2 $\pm$ 2.3&A\textsubscript{g}  \\\hline
    \end{tabular}
    \caption{\justifying Parameters extracted from multi-Lorentzian fitting of the room-temperature Raman spectrum of Ta\textsubscript{2}NiSe\textsubscript{5}. The table lists the phonon mode frequencies, their corresponding full width at half maximum (FWHM), and the assigned symmetry modes.}
    \label{tab:TS1}
\end{table}

\section{Experimental Setup}
\begin{figure}[h]
  \centering
  \includegraphics[width=0.85\textwidth]{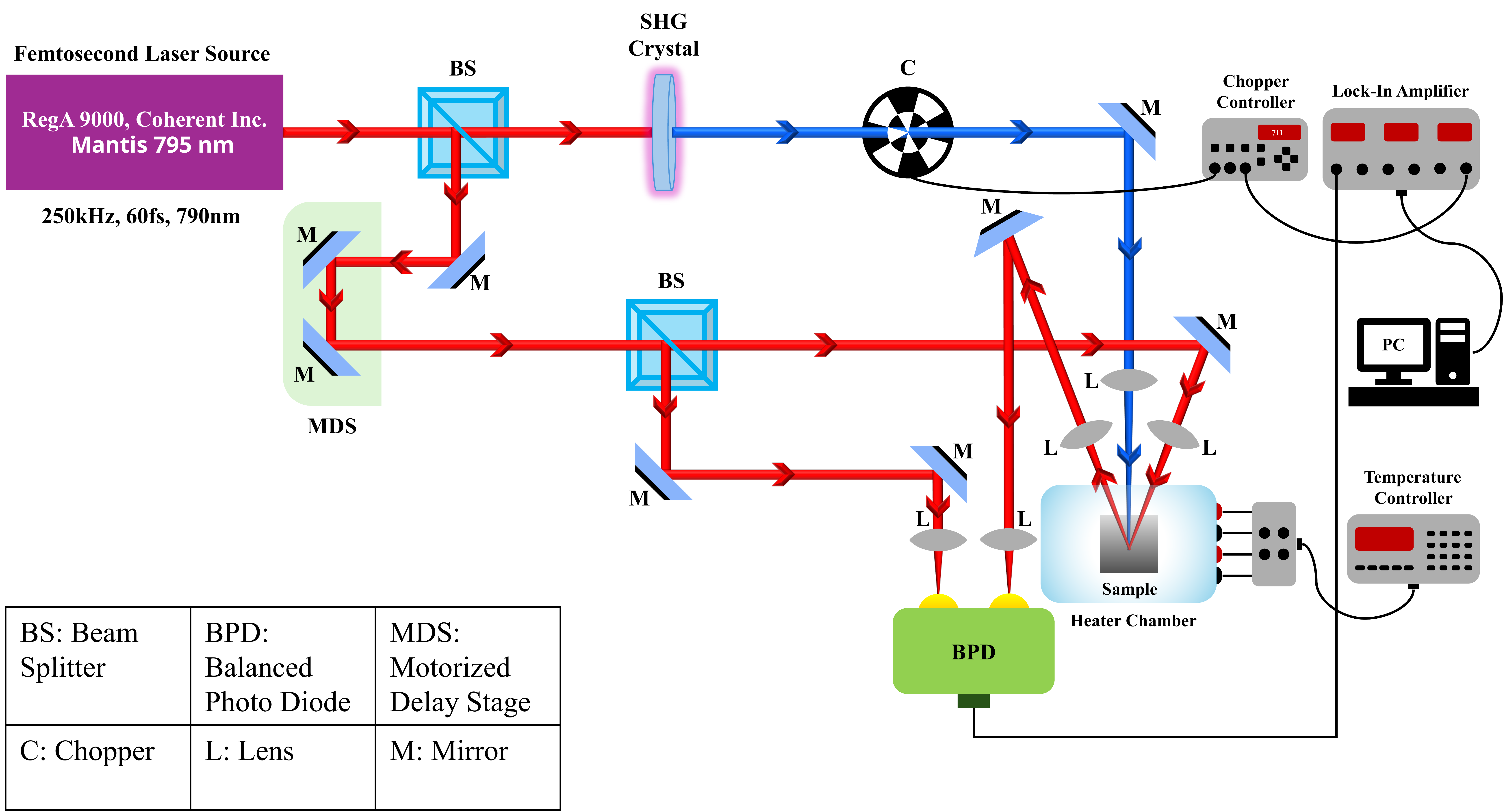}
  \caption{\justifying Schematic representation of the experimental setup for non-degenerate pump-probe spectroscopy performed in reflection geometry.}
  \label{fig:S4}
\end{figure}
Temperature-dependent non-degenerate pump-probe reflectivity measurements are performed to investigate the ultrafast carrier and phonon dynamics in the single crystal of  Ta\textsubscript{2}NiSe\textsubscript{5}. The experiments are conducted in a reflection geometry, as illustrated in Fig.\ref{fig:S4}. The laser pulses are obtained from a femtosecond regenerative amplifier (RegA 9050, Coherent Inc.) operating at a central wavelength of 790~nm with a repetition rate of 250~kHz. Each pulse has a duration of 60~fs and an energy of $6.4~\mu$J. The output beam is split into two parts in an 80:20 ratio. The higher-intensity portion is frequency-doubled to 390~nm using a type II $\beta$-barium borate (BBO) crystal and serves as the pump beam, while the remaining 790~nm beam is used as the probe. A computer-controlled motorized delay stage is introduced in the probe path to vary the temporal delay between the pump and probe pulses at the sample position. Both beams are spatially overlapped on the sample surface, with the pump incident near normal and the probe at a small angle. The sample is mounted in a temperature-controlled closed chamber under a dry nitrogen atmosphere to minimize environmental effects. Temperature control is provided by a home-built resistive heater consisting of a nichrome wire uniformly wound on a copper element to ensure efficient and homogeneous heat distribution, with the assembly electrically insulated and mechanically supported using a Teflon mount. The sample temperature is monitored by a calibrated temperature sensor (PT100) placed in close proximity to the sample and regulated using a CryoCon 32B temperature controller, providing stable temperature control within $\pm1$~K over the measurement range of 294~K to 350~K. The reflected probe signal, modulated by the pump excitation, is detected using a home-built balanced photodiode. To improve the signal-to-noise ratio, we use phase-sensitive lock-in detection, which is synchronized with an optical chopper modulating the pump beam at 711~Hz. The pump and probe beam spot diameters (FWHM) at the sample are approximately $59~\mu$m and $54~\mu$m, respectively. For temperature-dependent measurements, the pump and probe fluences have been fixed at $83~\mu\mathrm{J/cm^2}$  and $7~\mu\mathrm{J/cm^2}$, respectively. Care has been taken to ensure that no measurable sample degradation occurred within the chosen temperature or fluence ranges. The temporal resolution of the experiment has been determined by the cross-correlation of the pump and probe pulses in a BBO crystal, yielding an instrument response function of approximately 124~fs, corresponding to an actual pulse width of $\sim90$~fs assuming a sech\textsuperscript{2} pulse shape.

\section{Data Analysis}
\subsection{Temperature-Dependent Data}
\begin{figure}[h]
  \centering
  \includegraphics[width=\textwidth]{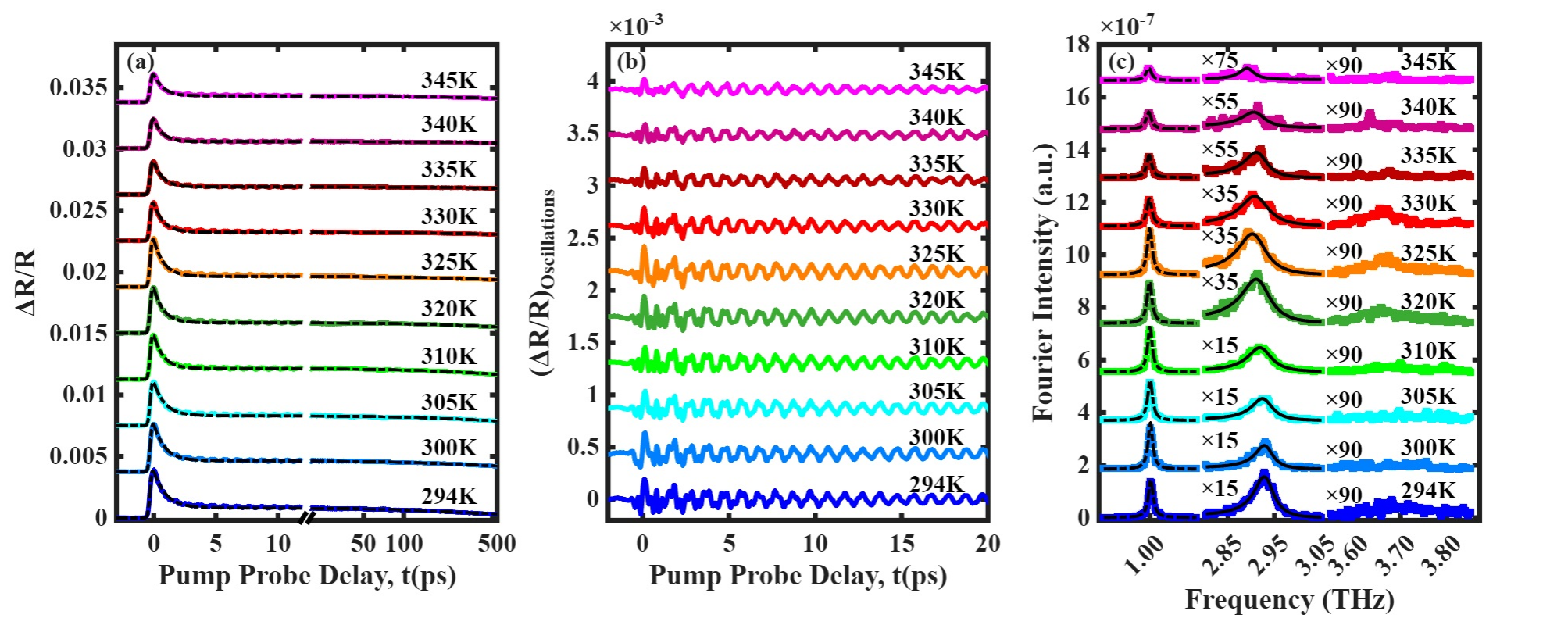}
  \caption{\justifying (a) Temperature-dependent time-resolved $\Delta R/R$ signals of TNSe at all temperatures (colored solid lines). The black dash-dotted lines represent fits to the data using Eq.(1) in the main manuscript. The x-axis is in linear scale till 15~ps, and after that, it is shown in log scale to view the entire data and fit with clarity. (b) Coherent oscillations extracted after subtracting the fits in (a) from the measured data. (c) Fourier intensity (FFTs) of the oscillations showing coherent optical phonon modes at 1~THz, 2.9~THz, and 3.7~THz (colored solid lines). The first two modes are fitted with Lorentzian and Fano line shapes, respectively, as shown by dash-dotted and solid black lines. Traces are vertically offset for better visualization.}
  \label{fig:S5}
\end{figure}

Fig.\ref{fig:S5} presents the complete set of temperature-dependent time-resolved $\Delta R/R$ measurements acquired in this study, covering all measured temperatures between 294~K and 350~K. In the main text, data traces at selected temperatures are shown for clarity, whereas the full dataset is provided here for completeness and transparency. The data shown in Fig.\ref{fig:S5} are analyzed following the same procedures described in the main text, including background fitting, extraction of coherent phonon oscillations, and Fourier analysis.

\subsection{Variation of A\textsubscript{0} and $\tau_r$ with Temperature}
\begin{figure}[h]
  \centering
  \includegraphics[width=\textwidth]{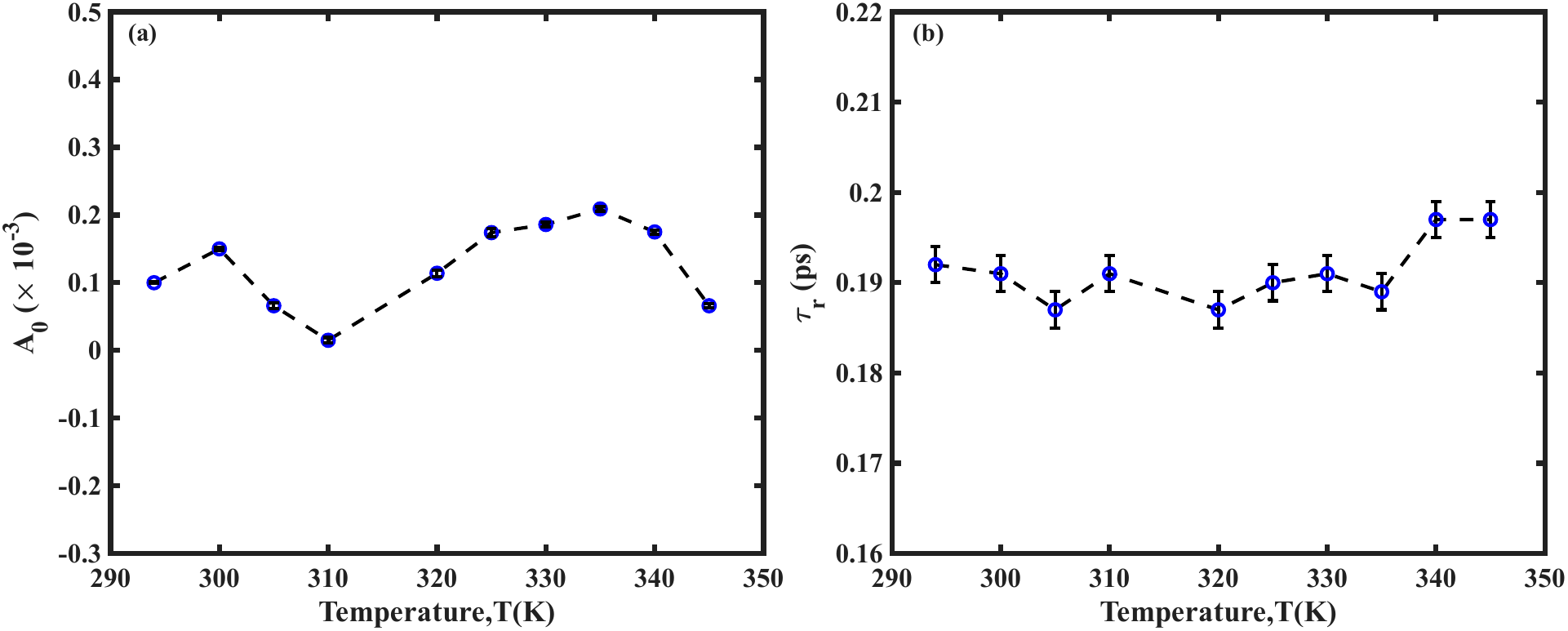}
  \caption{\justifying Temperature dependence of (a) long-lived amplitude A\textsubscript{0} and (b) rise time $\tau_r$ extracted from fitting of transient reflectivity data using Eq.(1) (in the main manuscript).}
  \label{fig:S6}
\end{figure}

For completeness, the temperature dependence of the long-lived amplitude (A\textsubscript{0}) and the rise time ($\tau_r$) obtained from the bi-exponential fitting using Eq.(1) (see main manuscript) are presented in Fig.\ref{fig:S6}. Both parameters remain nearly constant across the measured temperature range, indicating that the overall relaxation dynamics and baseline contribution are largely unaffected by temperature.

\subsection{Thermal-diffusivity Calculation and Estimation of Pump-induced Transient Heating}
Thermal diffusivity, $\alpha$ is calculated as

\begin{equation}
    \alpha = \frac{\kappa_c}{\rho\,C_p},
    \label{eq:alpha}
\end{equation}
Where, $\kappa_c$ is the thermal conductivity along the $c$-direction, 
$\rho$ is the mass density, and $C_p$ is the specific heat capacity at constant pressure.

For Ta\textsubscript{2}NiSe\textsubscript{5}, the in-plane thermal conductivity reported by \cite{zhang2021thermal} is $\kappa_c\approx20$ Wm\textsuperscript{-1}K\textsuperscript{-1} at 300~K.  
The density is $\rho=7.72\times 10^3$~kg/m\textsuperscript{3} \cite{Villars2023crystal} and the specific heat capacity at 300~K is $C_p\approx196$ JK\textsuperscript{-1}mol\textsuperscript{-1} \cite{haque2024terahertz, lu2017zero}. Using the molar mass of Ta\textsubscript{2}NiSe\textsubscript{5}, M = 815.44~gmol\textsuperscript{-1}, the corresponding volumetric heat capacity is given by

\begin{equation}
    C_V=\rho\left(\frac{C_p}{M}\right)\approx1.86\times10^6~\mathrm{J\,m^{-3}K^{-1}}
\end{equation}

Substituting these values into Eq.\ref{eq:alpha}, we get
\begin{equation}
    \alpha = 1.3\times10^{-5}\,\text{m}^2\,\text{s}^{-1}
\end{equation}

The optical penetration depth of TNSe at 1.57~eV (800~nm) and 2~eV has been reported to be approximately 55~nm in Ref. \cite{liu2021photoinduced} and approximately 30 nm in Ref. \cite{golez2022unveiling}, respectively. Since the absorption coefficient generally increases with photon energy above the optical gap, the penetration depth at the 3.14~eV (400~nm) pump wavelength is expected to be smaller. Accordingly, we estimate a pump penetration depth of approximately 20-25~nm for the present experiment.

Using this penetration depth, the characteristic thermal diffusion time, $t_{diff} = L^2/\alpha$, is estimated to be $\sim$30-50~ps for an optical penetration depth of 20-25~nm. Since the pulse separation at the repetition rate of 250~kHz is 4~$\mu$s, which is several orders of magnitude longer than $t_{diff}$, cumulative heating between successive laser pulses is expected to be negligible.

The same thermal parameters can also be used to estimate the transient lattice temperature rise induced by a single pump pulse. For an incident pump fluence of F = $83~\mu\mathrm{J/cm^2}$ with a measured reflectivity of $R=0.191$ at 400~nm, the absorbed energy within the probe sampling depth is estimated as
\begin{equation}
    E_{\mathrm{abs}} = F(1-R)\left(1-e^{-d/\delta_{\mathrm{pump}}}\right)
\end{equation}

where $\delta_{pump}=25$~nm is the estimated pump penetration depth and d = 55~nm is the probe penetration depth at 1.57 eV \cite{liu2021photoinduced}. The transient temperature rise is then obtained by equating the absorbed energy to the thermal energy required to heat the probe-sampled volume,
\begin{equation} 
    E_{\mathrm{abs}} = V\int_{T_0}^{T_f}n_{mol}C_p(T)\,dT
\end{equation}

where $V=A_{pump}d$ is the probe-sampled volume, $n_{mol}=\rho/M$ is the molar density and $C_p(T)$ is the temperature-dependent specific heat taken from Refs. \cite{haque2024terahertz,lu2017zero}. The integral is evaluated numerically to obtain the final lattice temperature, $T_f$, for each initial temperature, $T_0$.

For the lowest measurement temperature ($T_0=294$~K), the calculated transient temperature rise, $\Delta T$ is approximately 6~K, corresponding to a maximum transient temperature of approximately 300~K. Across the investigated temperature range (294-345~K), the transient temperature rise varies only weakly because of the modest temperature dependence of $C_p(T)$. These estimates confirm that the employed pump fluence produces only a small transient lattice heating and that the measurements were performed in the perturbative regime, well below the fluence required to photoinduce the high-temperature phase reported in previous high-fluence studies.

\subsection{Effect of Fitting Window on $\tau_2$}
\begin{figure}[h]
  \centering
  \includegraphics[width=\textwidth]{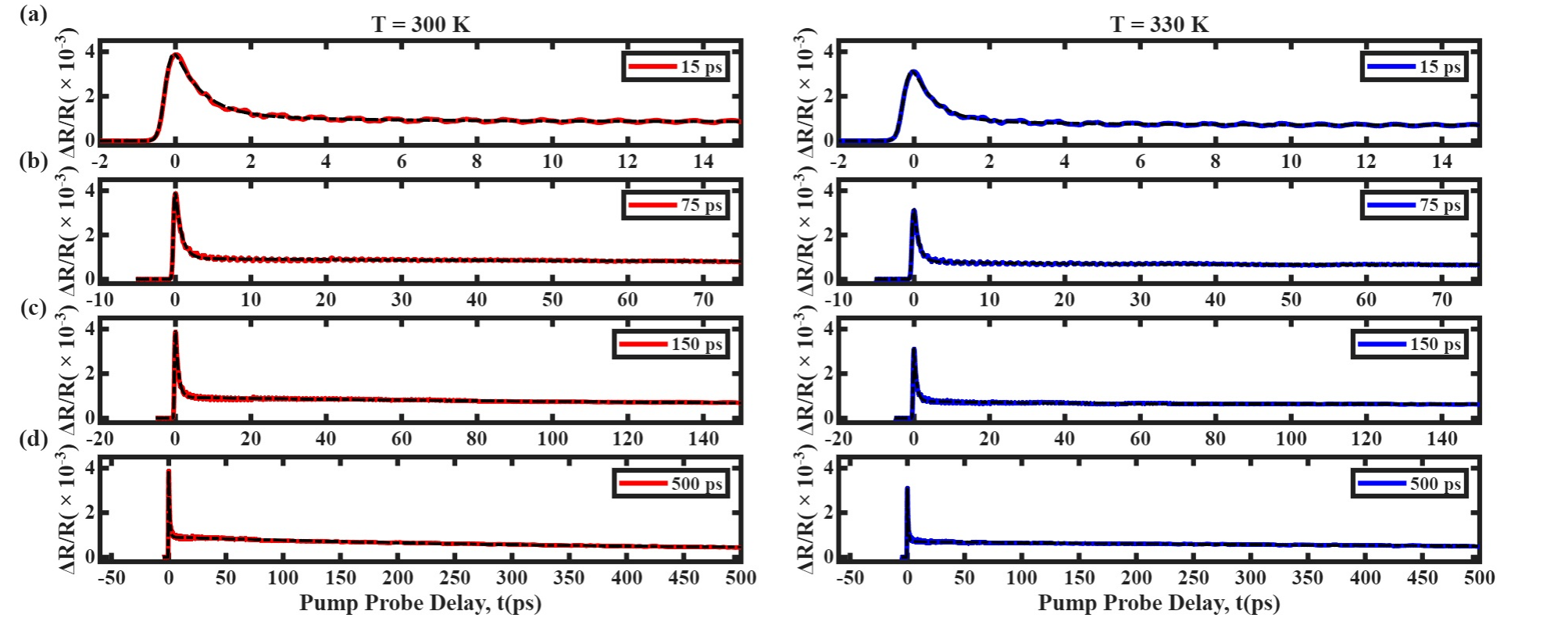}
  \caption{\justifying Transient reflectivity data at 300~K (left column) and 330~K (right column), with experimental data (solid lines) fitted over different upper time limits (dash-dotted lines for fit). Each row corresponds to a common fitting window: (a) 15~ps, (b) 75~ps, (c) 150~ps, and (d) 500~ps, illustrating the effect of the fitting range on the extracted relaxation time constant $\tau_2$. The corresponding fit parameters are summarized in Table TS2.}
  \label{fig:S7}
\end{figure}

To examine the dependence of the extracted relaxation time $\tau_2$ on the fitting window, the transient reflectivity data at 300~K and 330~K were re-fitted using different upper time limits of 15~ps, 75~ps, 150~ps, and 500~ps. Fig.\ref{fig:S7} displays representative fits of the data at 300~K and 330~K, respectively, for each fitting range. The corresponding fitting parameters, including $\tau_1$, $\tau_2$ and the goodness of fit parameter $R_s^2$, are summarized in Table \ref{tab:TS2}. For both temperatures, $\tau_2$ increases monotonically with the extension of the fitting window while maintaining good fit quality across all fitting ranges. This observation confirms that the inclusion of longer temporal windows captures additional slow relaxation channels as mentioned in the main manuscript.

\clearpage
\begin{table}[h]
    \centering

    \textbf{300 K}\\[6pt]
    \begin{tabular}{|>{\centering\arraybackslash}p{0.22\linewidth}|
                    >{\centering\arraybackslash}p{0.2\linewidth}|
                    >{\centering\arraybackslash}p{0.2\linewidth}|
                    >{\centering\arraybackslash}p{0.2\linewidth}|}
        \hline
        \textbf{Time Window (in ps)} & \textbf{$\tau_1$ (ps)} & \textbf{$\tau_2$ (ps)} & \textbf{$R_s^2$}\\
        \hline
        15  & 0.73 $\pm$ 0.02& 5.03   $\pm$ 1.55& 0.996\\\hline
        75  & 0.81 $\pm$ 0.01& 77.86  $\pm$ 19.60& 0.991\\\hline
        150 & 0.81 $\pm$ 0.01& 190.70 $\pm$ 15.06& 0.991\\\hline
        500 & 0.83 $\pm$ 0.01& 368.80 $\pm$ 3.38& 0.992\\\hline
    \end{tabular}

    \vspace{10pt}

    \textbf{330 K}\\[6pt]
    \begin{tabular}{|>{\centering\arraybackslash}p{0.22\linewidth}|
                    >{\centering\arraybackslash}p{0.2\linewidth}|
                    >{\centering\arraybackslash}p{0.2\linewidth}|
                    >{\centering\arraybackslash}p{0.2\linewidth}|}
        \hline
        \textbf{Time Window (in ps)} & \textbf{$\tau_1$ (ps)} & \textbf{$\tau_2$ (ps)} & \textbf{$R_s^2$}\\
        \hline
        15  & 0.58 $\pm$ 0.02& 3.84   $\pm$ 0.90& 0.995\\\hline
        75  & 0.64 $\pm$ 0.01& 17.26  $\pm$ 1.13& 0.990\\\hline
        150 & 0.66 $\pm$ 0.01& 34.31  $\pm$ 1.12& 0.989\\\hline
        500 & 0.75 $\pm$ 0.01& 600.00 $\pm$ 18.72& 0.982\\\hline
    \end{tabular}

    \caption{\justifying Fitting parameters obtained from transient reflectivity data fit at 300~K and 330~K for different upper time limits (15~ps, 75~ps, 150~ps, and 500~ps).}
    \label{tab:TS2}
\end{table}

\subsection{Fano Line Shape Analysis of the CP2 Mode at 2.9 THz}
\begin{figure}[h]
  \centering
  \includegraphics[width=\textwidth]{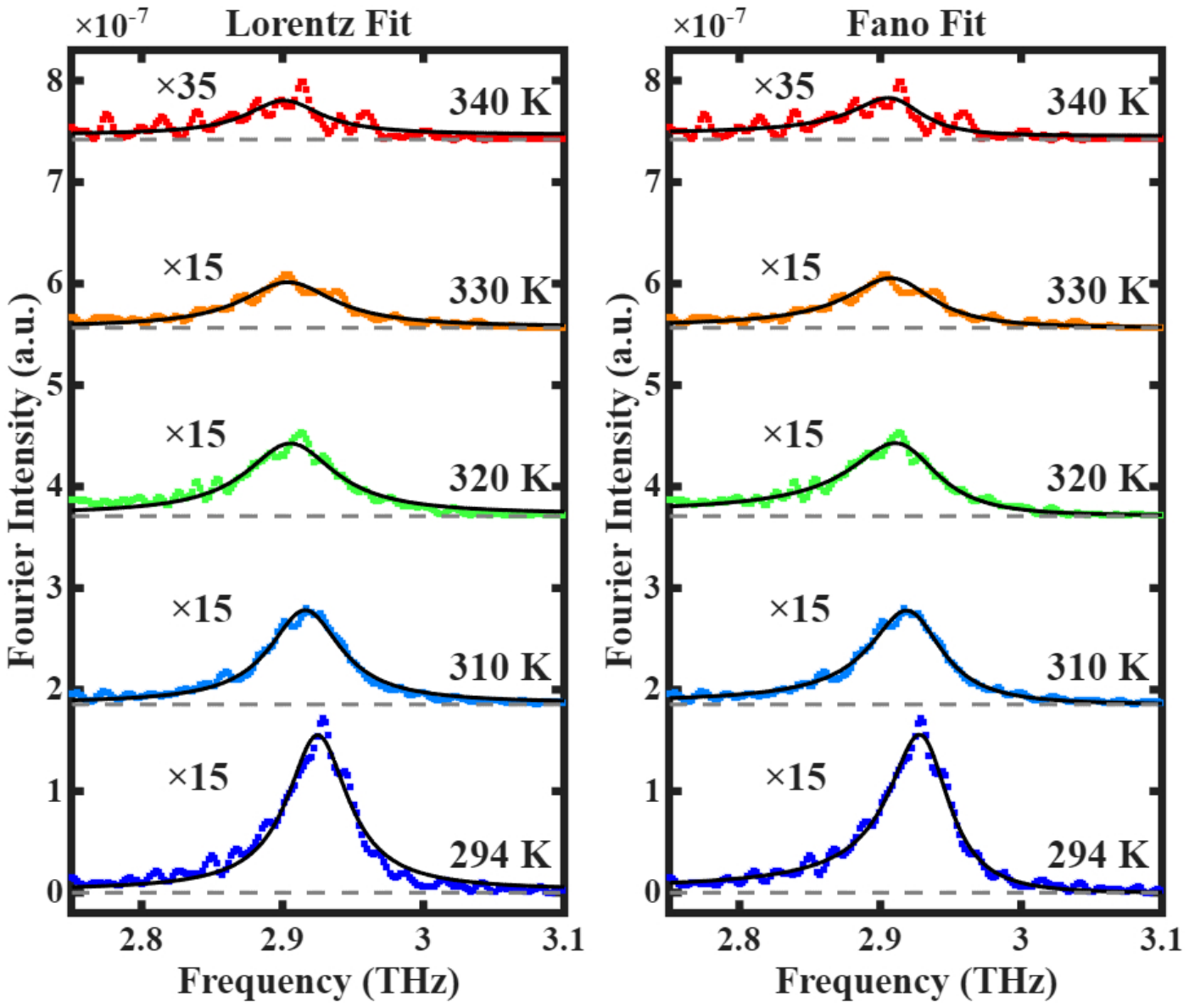}
    \captionof{figure}{\justifying A comparison between Lorentzian (left panel) and Fano (right panel) line-shape fits to the 2.9~THz coherent phonon mode at different temperatures is shown. The colored solid curves represent the experimental FFT data, while the black solid lines correspond to the respective fits in each panel. Each spectrum is vertically offset for clarity, with scaling factors indicated. Temperature values are labeled alongside each curve.}
    \label{fig:S8}
\end{figure}

We first attempted to describe the Fourier spectra using a conventional Lorentzian line shape. While the Lorentzian fit captures the overall spectral profile, a more detailed comparison reveals that the Fano line shape provides a better description at lower temperatures, where a clear asymmetry is more pronounced on the low-frequency (left) side of the peak. At higher temperatures, the two fits become nearly indistinguishable within the experimental resolution. To quantitatively assess the fitting quality, we compare the coefficients of determination ($R^2$), which are consistently higher for the Fano model across all measured temperatures, as summarized in Table \ref{tab:TS3}. Representative spectra and fits at selected temperatures are shown in Fig.\ref{fig:S8}. Based on this analysis and in line with prior reports \cite{werdehausen2018coherent,kim2021direct,liu2021photoinduced,arora2025ultrafast}, we adopt the Fano line shape for all subsequent analyses. A detailed physical interpretation of the Fano profile in terms of the coupling between the discrete phonon mode and the electronic continuum is discussed in the main text.

\begin{table}[h]
\centering
\begin{tabular}{|>{\centering\arraybackslash}p{0.3\linewidth}|>{\centering\arraybackslash}p{0.3\linewidth}|>{\centering\arraybackslash}p{0.3\linewidth}|}
    \hline
    \textbf{Parameter} & \textbf{Lorentzian} & \textbf{Fano} \\
    \hline
    \multicolumn{3}{|c|}{\textbf{294 K}} \\\hline
    Frequency (THz) & 2.924 $\pm$ 0.001 & 2.931 $\pm$ 0.001 \\\hline
    Linewidth (THz) & 0.053 $\pm$ 0.001 & 0.052 $\pm$ 0.001 \\\hline
    $R_s^2$         & 0.963 & 0.985 \\\hline

    \multicolumn{3}{|c|}{\textbf{310 K}} \\\hline
    Frequency (THz) & 2.916 $\pm$ 0.001 & 2.921 $\pm$ 0.001 \\\hline
    Linewidth (THz) & 0.062 $\pm$ 0.001 & 0.062 $\pm$ 0.001 \\\hline
    $R_s^2$         & 0.977 & 0.988 \\\hline

    \multicolumn{3}{|c|}{\textbf{320 K}} \\\hline
    Frequency (THz) & 2.906 $\pm$ 0.001 & 2.916 $\pm$ 0.001 \\\hline
    Linewidth (THz) & 0.077 $\pm$ 0.002 & 0.076 $\pm$ 0.001 \\\hline
    $R_s^2$         & 0.937 & 0.973 \\\hline

    \multicolumn{3}{|c|}{\textbf{330 K}} \\\hline
    Frequency (THz) & 2.904 $\pm$ 0.001 & 2.911 $\pm$ 0.001 \\\hline
    Linewidth (THz) & 0.080 $\pm$ 0.002 & 0.074 $\pm$ 0.002 \\\hline
    $R_s^2$         & 0.931 & 0.941 \\\hline

    \multicolumn{3}{|c|}{\textbf{340 K}} \\\hline
    Frequency (THz) & 2.901 $\pm$ 0.002 & 2.910 $\pm$ 0.002 \\\hline
    Linewidth (THz) & 0.065 $\pm$ 0.006 & 0.061 $\pm$ 0.005 \\\hline
    $R_s^2$         & 0.607 & 0.682 \\\hline
\end{tabular}
\caption{\justifying Comparison of Lorentzian and Fano fits to the Fourier spectrum of the 2.9~THz mode at different temperatures, showing the extracted frequency, linewidth, and coefficient of determination ($R_s^2$).}
\label{tab:TS3}
\end{table}
\clearpage

\subsection{Schematic Atomic Motions of Different Coherent Phonon Modes}
In order to provide a microscopic understanding of the observed coherent phonon oscillations, we illustrate the atomic displacements associated with the three dominant modes: 1~THz, 2.9~THz, and 3.7~THz. These visualizations help in identifying the nature of each mode and their contributions to the lattice dynamics.

\begin{figure}[h]
  \centering
  \begin{minipage}{0.48\textwidth}
        \centering
        \includegraphics[width=\linewidth]{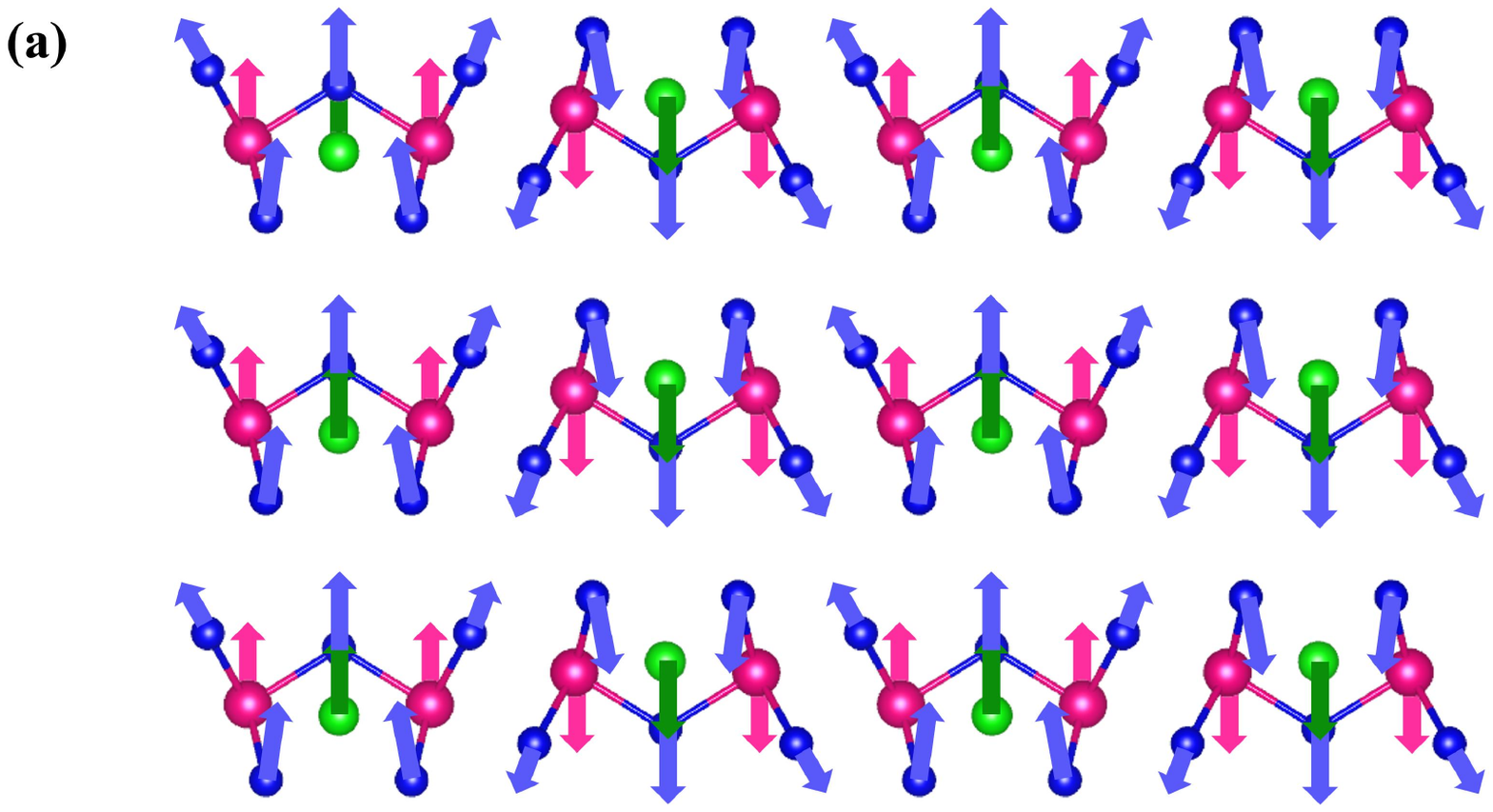}
    \end{minipage}
    \hfill
    \begin{minipage}{0.48\textwidth}
        \centering
        \includegraphics[width=\linewidth]{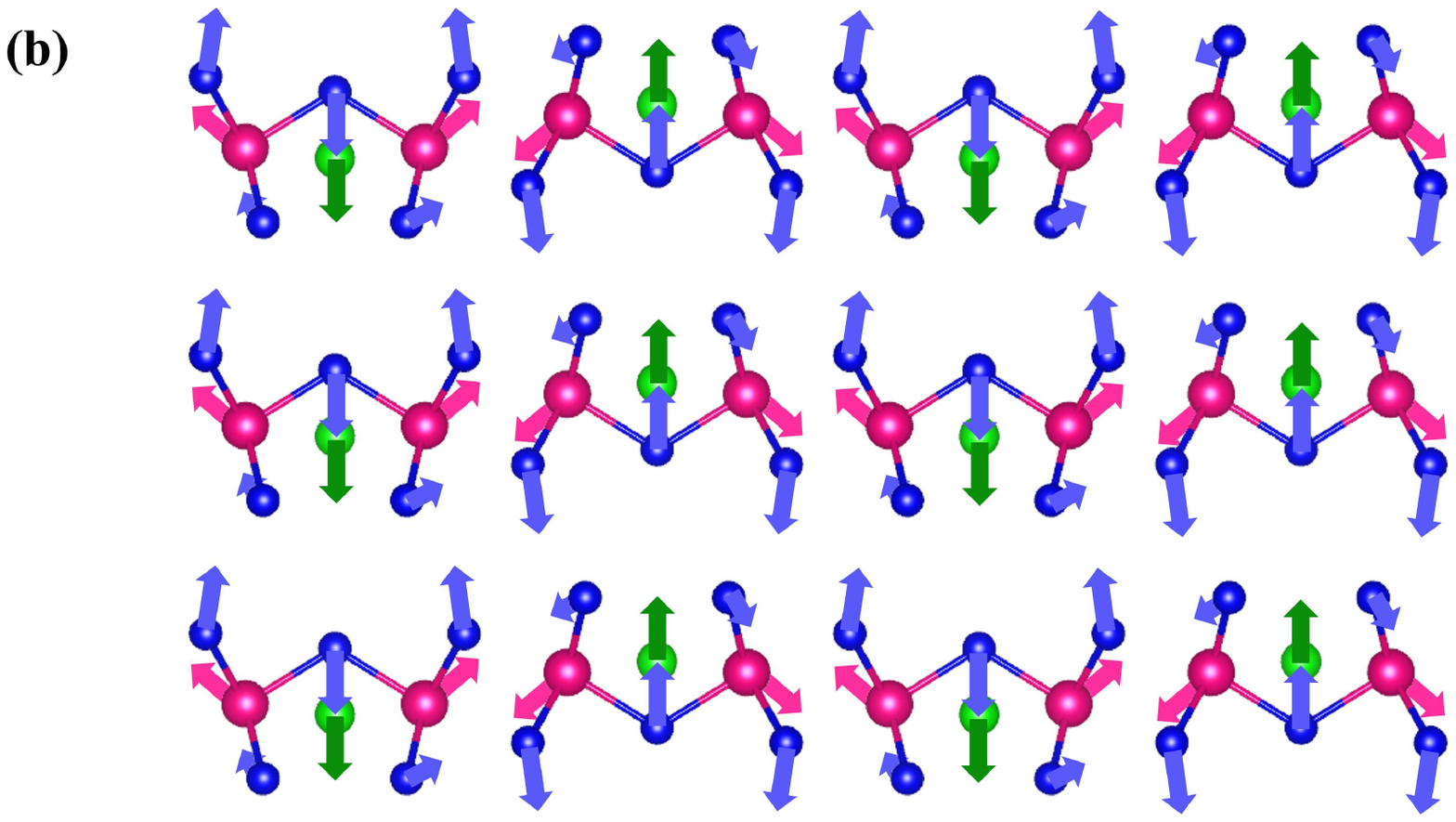}
    \end{minipage}

    \vspace{0.6em}

    \begin{minipage}{0.6\textwidth}
        \centering
        \includegraphics[width=\linewidth]{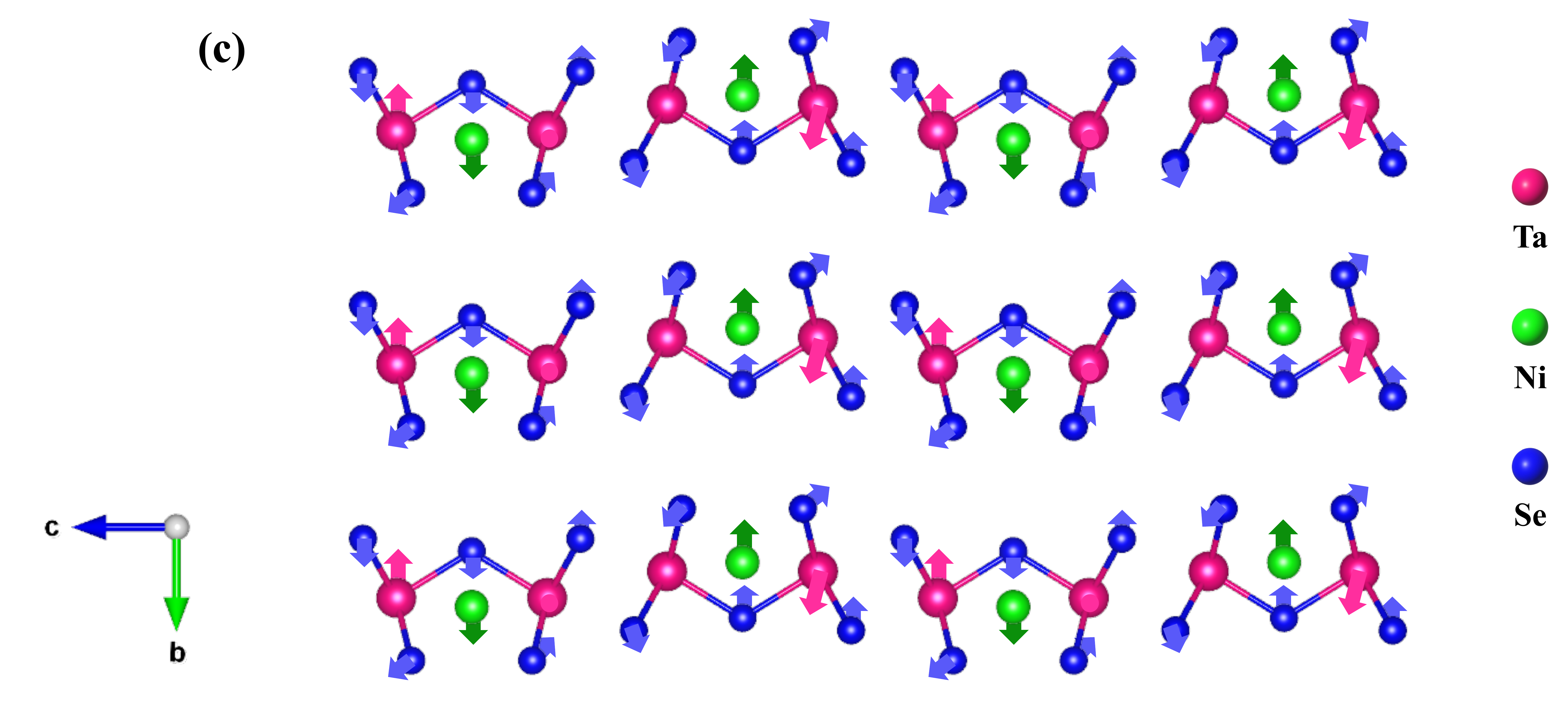}
    \end{minipage}
  \caption{\justifying Schematic representation of the atomic displacement patterns for the coherent phonon modes (CP1-CP3) in Ta$_2$NiSe$_5$. 
    (a) CP1 corresponds to the 1.0~THz coupled mode involving in-phase Ni-Ta motion with opposite Se displacement; 
   (b)  CP2 involves out-of-plane Se vibrations perpendicular to the Ta-Ni-Se layers; 
    (c) CP3 represents in-plane mixed Ta-Se vibrations within the layers.
    Arrows indicate the relative direction and phase of atomic motion.}
  \label{fig:S9}
\end{figure}

\section{Exciton-Phonon Bottleneck Model}

To describe the slow relaxation dynamics, we adopt a phenomenological exciton-phonon bottleneck model analogous to the Rothwarf-Taylor (RT) framework \cite{rothwarf1967measurement,kabanov2005kinetics,werdehausen2018photo}. Following photoexcitation, the excitonic state is strongly perturbed, producing photoexcited carriers and a non-equilibrium phonon population. The subsequent recombination and recovery of excitonic correlations are accompanied by energy transfer to the lattice. We describe the coupled populations of photoexcited electronic excitations ($n$) and non-equilibrium phonons ($N_{ph}$) by

\begin{equation}
    \frac{dn}{dt} = -R n^2 + \eta N_{ph}; \quad
    \frac{dN_{ph}}{dt} = \frac{1}{2} R n^2 - \frac{N_{ph}}{\tau_{ph}},
\end{equation}

where, $R$ is the effective recombination rate, $\eta$ accounts for phonon-assisted re-excitation and $\tau_{ph}$ is the phonon relaxation time. This simplified form neglects explicit source terms and equilibrium populations, focusing on post-excitation dynamics.

The bottleneck model employed here is not a microscopic model specific to TNSe. However, its central ingredient, strong coupling between excitonic and lattice degrees of freedom, is well established in this material. In this context, previous studies have also proposed that the relevant excitonic states may be strongly dressed by lattice excitations. Larkin \textit{et al.} \cite{larkin2017giant} interpreted the giant excitonic Fano resonance in TNSe in terms of spatially extended exciton-polaron complexes arising from strong exciton-phonon coupling, while Jiang \textit{et al.} \cite{jiang2024multiple} showed theoretically that exciton-phonon coupling can stabilize lower-energy phonon-dressed excitonic states. Thus, an exciton-polaron-like state provides one possible microscopic picture of the strongly coupled exciton-lattice system. Our measurements, however, do not distinguish between bare excitonic and phonon-dressed excitonic states and we therefore retain the more general exciton-phonon bottleneck description.

Within this framework, the bottleneck arises because the non-equilibrium phonon population generated during relaxation can feed energy back into the electronic system, delaying the restoration of equilibrium excitonic correlations until the phonon population relaxes. The resulting slow recovery can therefore be governed by the lattice relaxation timescale rather than solely by the intrinsic electronic recombination rate. In TNSe, where excitonic correlations are strongly coupled to lattice distortions, this provides a natural phenomenological picture for the long-lived relaxation channel observed in our measurements. The fast component is associated with the initial carrier relaxation and recombination, whereas the slow component ($\sim280$-$600$~ps) reflects the recovery of the coupled excitonic-lattice system in the presence of a non-equilibrium phonon population.

\section{Pump Fluence-Dependence of Transient Relaxation Dynamics}
\begin{figure}[hb]
  \centering
  \includegraphics[width=\textwidth]{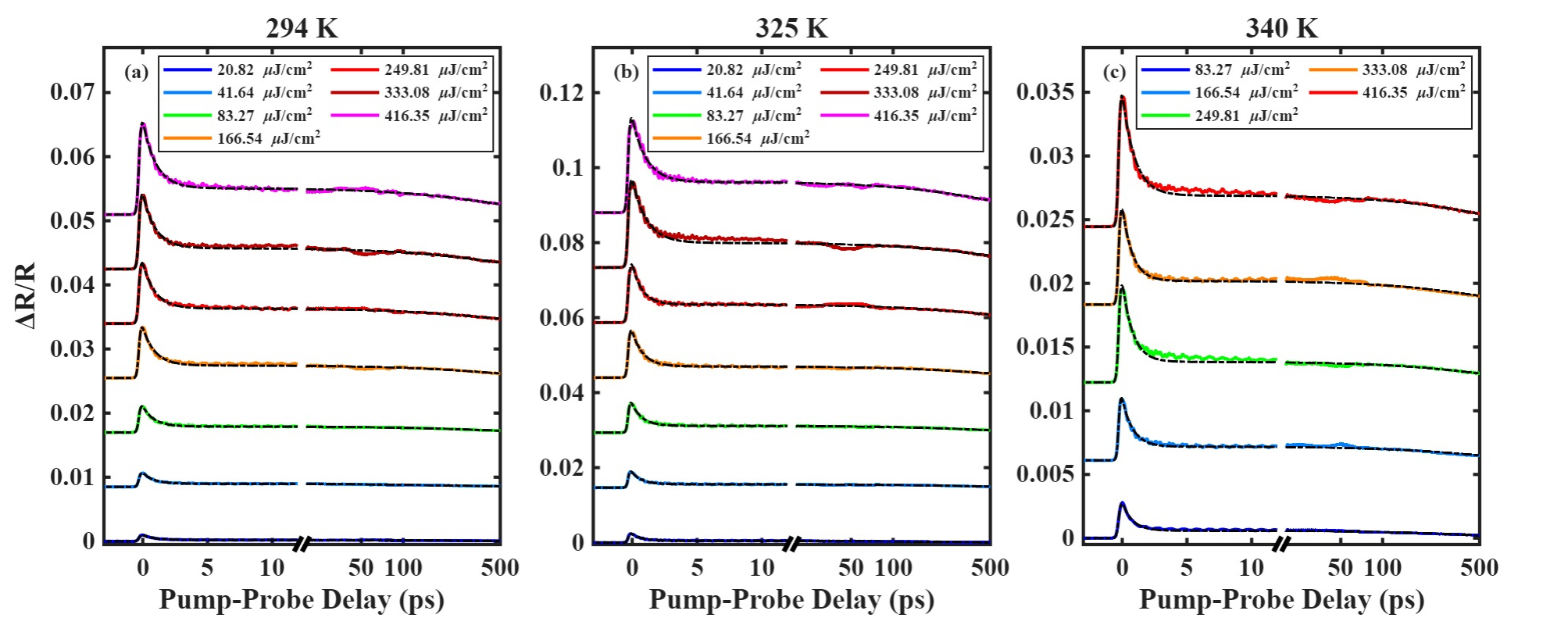}
  \caption{\justifying Pump-fluence-dependent transient reflectivity ($\Delta R/R$) signals of TNSe measured at (a) 294~K, (b) 325~K and (c) 340~K for different pump fluences (colored solid lines), shown up to a pump-probe delay of 500~ps. The black dash-dotted lines represent the fits to the experimental data using Eq. (1) of the main manuscript. The time axis is displayed on a linear scale up to 15~ps and on a logarithmic scale thereafter to clearly visualize the complete transient dynamics and the quality of the fits.}
  \label{fig:S10}
\end{figure}

\begin{figure}[h]
  \centering
  \includegraphics[width=\textwidth]{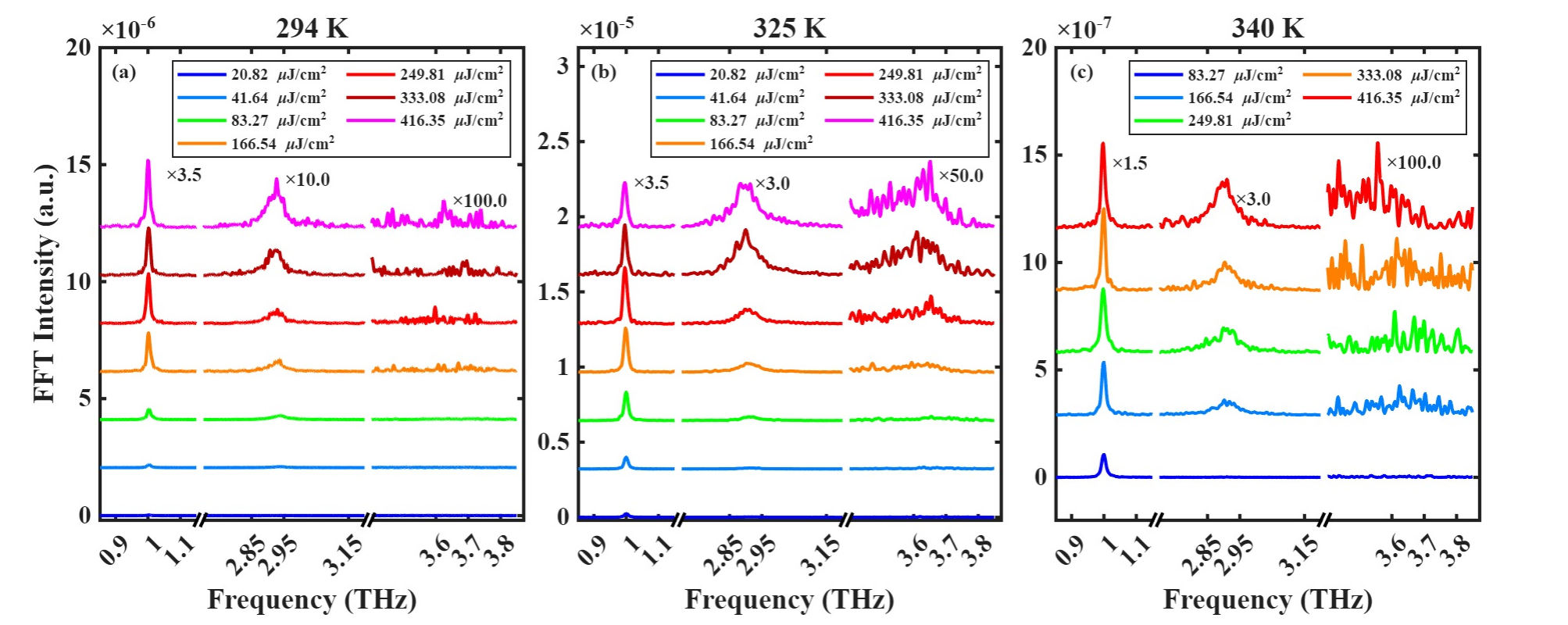}
  \caption{\justifying Coherent oscillations extracted from the pump-fluence-dependent transient reflectivity data after subtracting the fitted background shown in Fig. S10. The corresponding FFT spectra reveal the evolution of the coherent phonon modes with increasing pump fluence at (a) 294~K, (b) 325~K and (c) 340~K.}
  \label{fig:S11}
\end{figure}

\begin{figure}[h]
  \centering
  \includegraphics[width=0.8\textwidth]{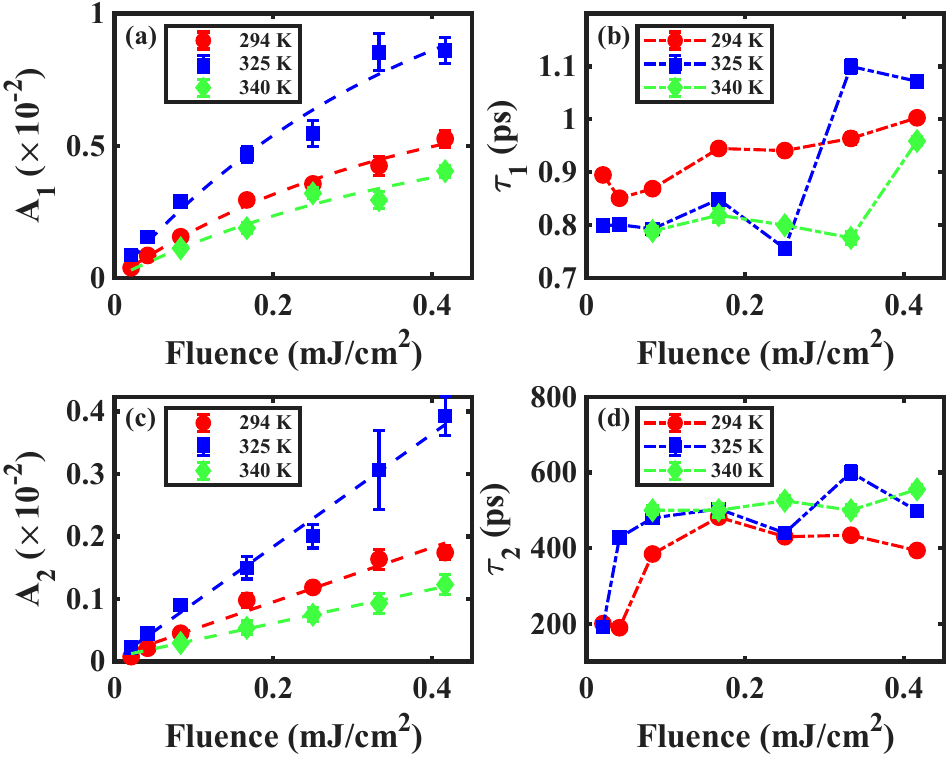}
  \caption{\justifying Pump-fluence dependence of the fitting parameters extracted from the bi-exponential fitting of the transient reflectivity data shown in Fig. S10 at 294~K (red circles), 325~K (blue squares) and 340~K (green diamonds). (a) Fast relaxation amplitude, A\textsubscript{1}, (b) fast relaxation time, $\tau_1$, (c) slow relaxation amplitude, A\textsubscript{2} and (d) slow relaxation time, $\tau_2$. The dashed and dash-dotted lines are guides to the eye.}
  \label{fig:S12}
\end{figure}

\begin{figure}[h]
  \centering
  \includegraphics[width=\textwidth]{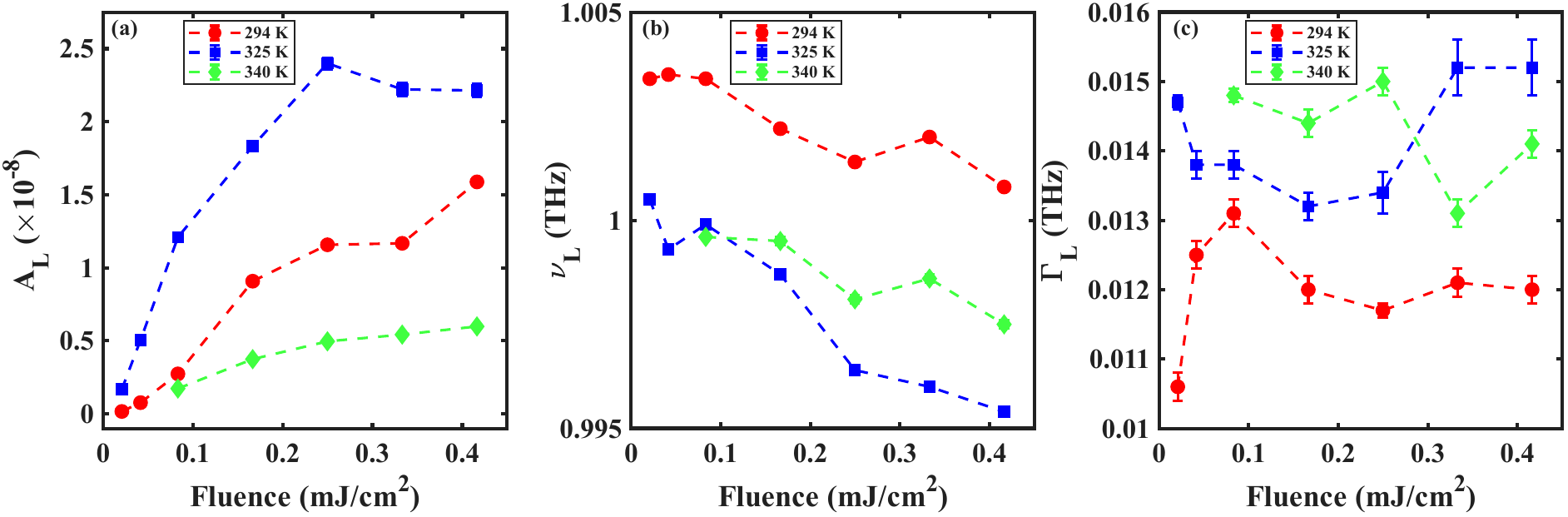}
  \caption{\justifying Pump-fluence dependence of the parameters of the dominant coherent phonon mode obtained from Lorentzian fits to the FFT spectra at 294~K (red circles), 325~K (blue squares) and 340~K (green diamonds). (a) FFT peak amplitude ($A_L$), (b) phonon frequency ($\nu_L$) and (c) linewidth ($\Gamma_L$) of the $\sim$1 THz mode as a function of pump fluence. The dashed lines are guides to the eye.}
  \label{fig:S13}
\end{figure}

To investigate the influence of excitation density on the relaxation dynamics of TNSe, we performed pump-fluence-dependent transient reflectivity ($\Delta R/R$) measurements at 294~K (below $T_c$), 325~K (near $T_c$) and 340~K (above $T_c$) with pump fluences ranging from $21~\mu\mathrm{J/cm^2}$ to $416~\mu\mathrm{J/cm^2}$ at a fixed probe fluence of $7~\mu\mathrm{J/cm^2}$. The transient reflectivity traces and their corresponding bi-exponential fits [using Eq. (1) of the main manuscript] are shown in Fig. \ref{fig:S10}.

The extracted fit parameters are summarized in Fig. \ref{fig:S12}. Both the fast (A\textsubscript{1}) and slow (A\textsubscript{2}) amplitudes increase monotonically with pump fluence across all temperatures, reflecting the larger photoexcited carrier density. For any given fluence, the amplitudes peak near $T_c$ (325~K) and are lower at 340~K than at 294~K, consistent with the temperature dependence discussed in the main manuscript. The fast relaxation time, $\tau_1$, displays a subtle, systematic increase with pump fluence at all temperatures. In contrast, the slow relaxation time, $\tau_2$, increases sharply up to $\sim$~80-100~$\mu\mathrm{J/cm^2}$ and approaches saturation at higher fluences. The fluence dependence of the relaxation times here provides additional insight into the two relaxation pathways. The gradual increase of $\tau_1$ with pump fluence indicates that the initial carrier relaxation becomes slower at higher excitation densities, consistent with density-dependent interactions and screening. In contrast, the rapid increase of $\tau_2$ at low fluence followed by saturation at higher fluence indicates that the long-lived relaxation is not simply proportional to the photoexcited carrier density. The saturation is consistent with a bottleneck in which, beyond a certain excitation density, the recovery becomes increasingly limited by the relaxation of the coupled electronic-lattice system. While these observations do not by themselves establish a microscopic origin for the bottleneck, they provide additional support for the phenomenological exciton-phonon bottleneck picture discussed in the main text.

To examine the coherent phonon response, Fourier transform (FFT) spectra were obtained from the oscillatory residuals after subtracting the non-oscillatory decay background [Figs. \ref{fig:S11}(a)-(c)]. The FFT spectra show that the frequencies of the dominant coherent phonon mode ($\sim$1~THz) remain unaffected over the entire investigated fluence range. Lorentzian fitting parameters for this mode are further detailed in Fig. \ref{fig:S13}. Overall, increasing the pump excitation density primarily scales the electronic amplitudes and relaxation times, leaving the coherent phonon spectrum qualitatively unchanged.

\clearpage

\section{Temperature Dependence of the Equilibrium Reflectivity at the Probe Energy}

\begin{figure}[h]
  \centering
  \includegraphics[width=0.8\textwidth]{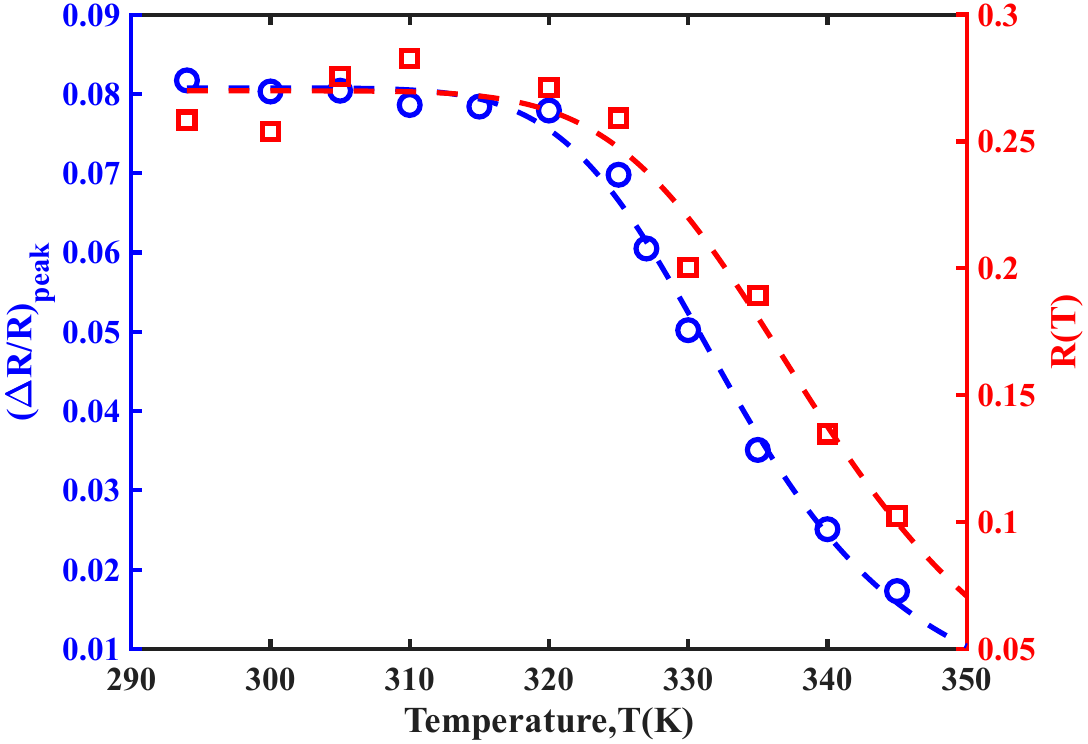}
  \caption{\justifying Temperature dependence of the peak transient reflectivity amplitude (left axis, blue) and the equilibrium reflectivity (right axis, red), measured at the probe photon energy of 1.57~eV. The dashed curves represent fits using Eq. (2) of the main manuscript.}
  \label{fig:S14}
\end{figure}

To examine the possible contribution of the temperature-dependent optical response to the transient-reflectivity amplitude, we compare in Fig. \ref{fig:S14} the equilibrium reflectivity $R(T)$ measured at the probe photon energy of 1.57~eV with the peak transient-reflectivity amplitude, $(\Delta R/R)_{peak}$ at $t=0$ ps. The two quantities exhibit a qualitatively similar temperature dependence over the investigated temperature range.

The 1.57~eV probe is close to interband transitions involving electronic states near the $\Gamma$-point of the Brillouin zone. Temperature-dependent changes in the transition energies and optical matrix elements can therefore modify the dielectric response and the sensitivity of the probe to a pump-induced perturbation. The similarity between $R(T)$ and $(\Delta R/R)_{peak}$ indicates that the evolving equilibrium optical response at the probe energy may contribute to the temperature dependence of the measured transient amplitude.

For comparison, $R(T)$ can also be described phenomenologically by the same temperature-dependent functional form used to analyze the transient amplitude. This comparison should not be interpreted as evidence that the equilibrium reflectivity follows Rothwarf-Taylor dynamics. Rather, it demonstrates that the temperature dependence of the transient amplitude is not uniquely determined by the photoexcited population. In the present single-color pump-probe measurement, the contribution from the temperature-dependent probe optical response cannot be quantitatively separated from the intrinsic temperature dependence of the photoinduced response. We therefore interpret the RT analysis of the transient amplitude as being consistent with, but not uniquely establishing, the evolution of the correlation gap.

\end{document}